\documentclass[acmtog,nonacm]{acmart}

\usepackage{booktabs} % For formal tables
\usepackage[table]{xcolor}
\usepackage{multirow}
\usepackage{wrapfig}
\usepackage{enumitem}
\usepackage[colorinlistoftodos]{todonotes}
\usepackage[linesnumbered,ruled,vlined]{algorithm2e}

\usepackage{amsthm}

\usepackage[ruled]{algorithm2e} % For algorithms

\SetAlFnt{\small}
\SetAlCapFnt{\small}
\SetAlCapNameFnt{\small}
\SetAlCapHSkip{0pt}

\begin{document}
\begin{teaserfigure}
  \includegraphics[width=\textwidth]{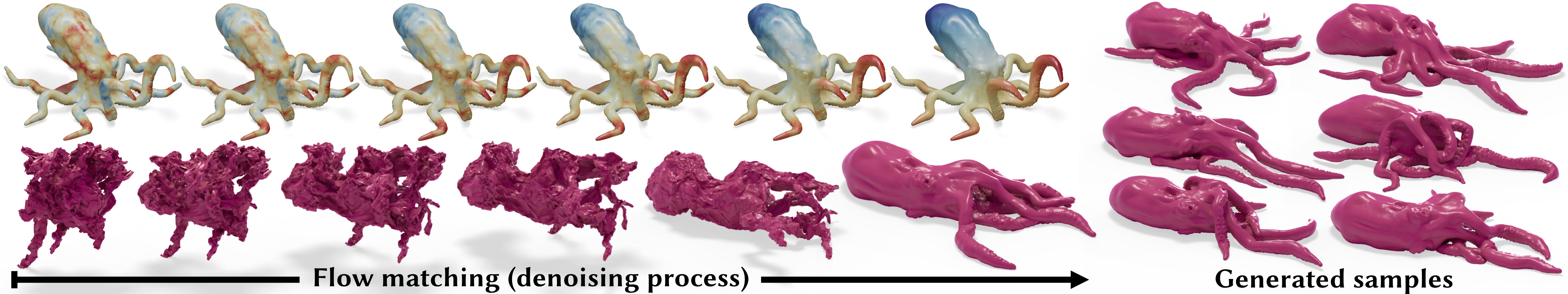}
  \caption{\textbf{Generating different deformations of a high-resolution mesh through triangulation-agnostic flow matching.} Top left: our method generates signals (visualized via colors) on meshes via the flow matching~\cite{lipman2022flow} paradigm's denoising process.  Bottom left: in this case the signals generated correspond to a  deformation of the mesh's vertices. The model was trained on a dataset of physical simulation of elastic equilibrium states, and thus produces various elastic rest poses of the 3D shape (right). The training set comprised of meshes in resolution of \emph{9k} faces, and through the triangulation agnosticism of our method we generate results at test time at a resolution of \emph{100k} faces.}
  \label{fig:teaser}
\end{teaserfigure}
% Title portion
\title{\matern{} Noise for Triangulation-Agnostic Flow Matching on Meshes}

% DO NOT ENTER AUTHOR INFORMATION FOR ANONYMOUS TECHNICAL PAPER SUBMISSIONS TO SIGGRAPH 2019!
%\author{Gang Zhou}
%\orcid{1234-5678-9012-3456}
%\affiliation{%
%  \institution{College of William and Mary}
%  \streetaddress{104 Jamestown Rd}
%  \city{Williamsburg}
%  \state{VA}
%  \postcode{23185}
%\email{gang_zhou@wm.edu}
%\affiliation{%
%  \institution{Inria Paris-Rocquencourt}
%\email{beranger@inria.fr}
%\affiliation{%
% \institution{Rajiv Gandhi University}
%\email{aprna_patel@rguhs.ac.in}
%\affiliation{%
%  \institution{Tsinghua University}
%  \streetaddress{30 Shuangqing Rd}
%  \country{China}
%}
%\email{chan0345@tsinghua.edu.cn}
%\author{Ting Yan}
%\affiliation{%
%  \institution{Eaton Innovation Center}
%  \city{Prague}
%  \country{Czech Republic}}
%\email{yanting02@gmail.com}
%\author{Tian He}
%\affiliation{%
%  \institution{University of Virginia}
%  \department{School of Engineering}
%  \city{Charlottesville}
%  \state{VA}
%  \postcode{22903}
%  \country{USA}
%}
%\affiliation{%
%  \institution{University of Minnesota}
%\email{tinghe@uva.edu}
%\author{Chengdu Huang}
%\author{John A. Stankovic}
%\author{Tarek F. Abdelzaher}
%\affiliation{%
%  \institution{University of Virginia}
%  \department{School of Engineering}
%  \city{Charlottesville}
%  \state{VA}
%  \postcode{22903}
%  \country{USA}
%}

%\renewcommand\shortauthors{Zhou, G. et al}

\begin{abstract}
This paper tackles the task of learning to generate signals over triangle meshes in a triangulation-agnostic manner, meaning the trained model can be applied to different meshes and triangulations effectively. Practically, the paper adapts the flow matching (FM) paradigm to a mesh-based, triangulation-agnostic setting. Theoretically, it proposes a specific noise distribution which is triangulation agnostic, to be used inside the FM model's denoising process. While  noise distributions are usually trivial to devise for, e.g., images, devising a triangulation-agnostic distribution proves to be a much more difficult task. We formulate a mathematical definition of triangulation agnosticism of distributions, via their spectrum. We then show that a discretization of a specific Gaussian random field called a \emph{\matern{} process} holds these desired properties, and provides a simple and efficient sampling algorithm. We use it as our noise model, and adapt FM to the triangulation-agnostic setting by using a state-of-the-art approach for learning signals on meshes in the gradient domain---PoissonNet---as the denoiser. We conduct experiments on elaborate tasks such as sampling elastic rest states, and generating poses of  humanoids. Our method is shown to be capable of producing highly realistic results for meshes of over one million triangles, significantly exceeding the state-of-the-art in quality and diversity.
\end{abstract}


\author{Tianshu Kuai}
% \orcid{0009-0007-9880-1293}
% \email{tianshu.kuai@umontreal.ca}
\affiliation{%
  \institution{Université de Montréal \& Mila}
  \country{Canada}
}
\author{Arman Maesumi}
% \orcid{0000-0001-7898-8061}
% \email{arman_maesumi@brown.edu}
\affiliation{%
  \institution{Brown University}
  \country{USA}
}
\author{Daniel Ritchie}
% \orcid{0000-0002-8253-0069}
% \email{daniel_ritchie@brown.edu}
\affiliation{%
  \institution{Brown University}
  \country{USA}
}
\author{Noam Aigerman}
% \orcid{0000-0002-9116-4662}
% \email{noam.aigerman@umontreal.ca}
\affiliation{%
  \institution{Université de Montréal \& Mila}
  \country{Canada}
}

\newcommand{\manifold}{\Omega}
\newcommand{\mesh}{\mathcal{M}}
\newcommand{\othermesh}{\mathcal{N}}
\newcommand{\mass}{\textbf{M}}
\newcommand{\identity}{\textbf{I}}

\newcommand{\lap}{\textbf{L}}
\newcommand{\real}{\mathbb{R}}
\newcommand{\zvec}{\mathbf{0}}
\newcommand{\ovec}{\mathbf{1}}
\newcommand{\evecmat}{\mathbf{\Phi}}
\newcommand{\evalmat}{\mathbf{\Lambda}}
\newcommand{\evec}{\boldsymbol{\phi}}

\newcommand{\covarmat}{\mathbf{\Sigma}}
\newcommand{\meanvec}{\mathbf{\mu}}

\newcommand{\brac}[1]{\left[#1\right]}
\newcommand{\tribrac}[1]{\left<#1\right>}
\newcommand{\variance}[1]{\text{Var}\brac{#1}}
\newcommand{\parr}[1]{\left(#1\right)}
\newcommand{\set}[1]{\left\{#1\right\}}
\newcommand{\bigo}[1]{\mathcal{O}\parr{#1}}
\newcommand{\mypar}[1]{%
\par
  \vspace{1ex}          % Add a little space above
  \noindent\textbf{#1}  % Bold and no indent
}

\newcommand{\lbo}{\mathcal{L}}
\newcommand{\signal}{\textbf{f}}
\newcommand{\coef}{\widehat{\signal}}
\newcommand{\dist}{\mathcal{D}}
\newcommand{\spectraldist}{\widehat{\dist}}
\newcommand{\normaldist}[2]{\mathcal{N}\left(#1,#2\right)}
\newcommand{\whitedist}{\dist_\text{white}}
\newcommand{\whitenoise}{\textbf{w}}
\newcommand{\spectralwhitenoise}{\widehat{\whitenoise}}
\newcommand{\spectralwhitedist}{\spectraldist_\text{white}}
\newcommand{\poissondist}{\dist_\text{\matern{}}}
\newcommand{\spectralpoissondist}{\spectraldist_\text{\matern{}}}
\newcommand{\matern}{Matérn}


% \setcopyright{cc}
% \setcctype{by}
% \acmJournal{TOG}
% \acmYear{2026} \acmVolume{45} \acmNumber{4} \acmArticle{72}
% \acmMonth{7} \acmDOI{10.1145/3811309}

\maketitle

\section{Introduction}

In recent years, there has been an unprecedented surge in the generative capabilities of machine learning models, enabled by novel techniques such as denoising diffusion models~\cite{ho2020denoising,song2020score} and flow matching~\cite{lipman2022flow}. These methods have been successfully applied to various  visual modalities, e.g.,  2D pixel images~\cite{yu2025pixeldit}, and 3D voxel grids~\cite{xiang2025structured}.

As of today, {triangle meshes} stand as a modality that has yet to obtain practical generative capabilities of the same quality.
In this context,  by ``generative capabilities for meshes'', we mean treating an \emph{existing} mesh as the domain---analogous to, e.g., the pixel grid of an image---and generating values over its vertices. For instance, in Figure~\ref{fig:teaser}, our method generates values representing  $(x,y,z)$ coordinates of the vertices of an octopus mesh, resulting in different deformations of the shape. 

As meshes are one of the most ubiquitous representations of 2-manifolds---used in, e.g., computer graphics, video games, engineering, and biomedical imaging---introducing such generative capabilities holds potential for impactful applications.

This work tackles one of the main roadblocks to a practical mesh-based generative approach, and devises a  generative framework that is \emph{triangulation agnostic}~\cite{aigerman2022neural,sharp2022diffusionnet,maesumi2025poissonnet}, meaning that the method can be applied to arbitrary triangulations of a 3D model and will exhibit near-identical behavior, regardless of the chosen triangulation (as long as it is well-behaved enough). 

Indeed, triangulation agnosticism is crucial for real-world applications involving meshes, as it provides two critical benefits:
\begin{enumerate}[label=\textbf{\arabic*)}, wide=0pt, nosep]
\item \textbf{Applicability to heterogeneous data.}
In contrast to images, meshes are irregular graphs representing arbitrary discretizations of an underlying 3D object, and thus mesh datasets contain models with varying triangulations. It has been shown that naive architectures are sensitive to spurious information in the triangulations of such training data, which leads to invalid results at test time~\cite{sharp2022diffusionnet} and has prompted the community to prefer architectures that behave agnostically under changes to the triangulation. 
\item \textbf{Efficiency on high-resolution meshes.}
The computations in triangulation-agnostic methods are less coupled to the triangulation and hence to its resolution, which leads to better performance, and allows triangulation-agnostic methods to benefit from efficient training (at lower resolutions) while still being applicable for inference on high-resolution meshes. Indeed, we demonstrate that our method attains state-of-the-art results, at a resolution that far exceeds prior generative methods for meshes.
\end{enumerate}
\begin{figure}[t]       
    \centering      \includegraphics[width=\linewidth]{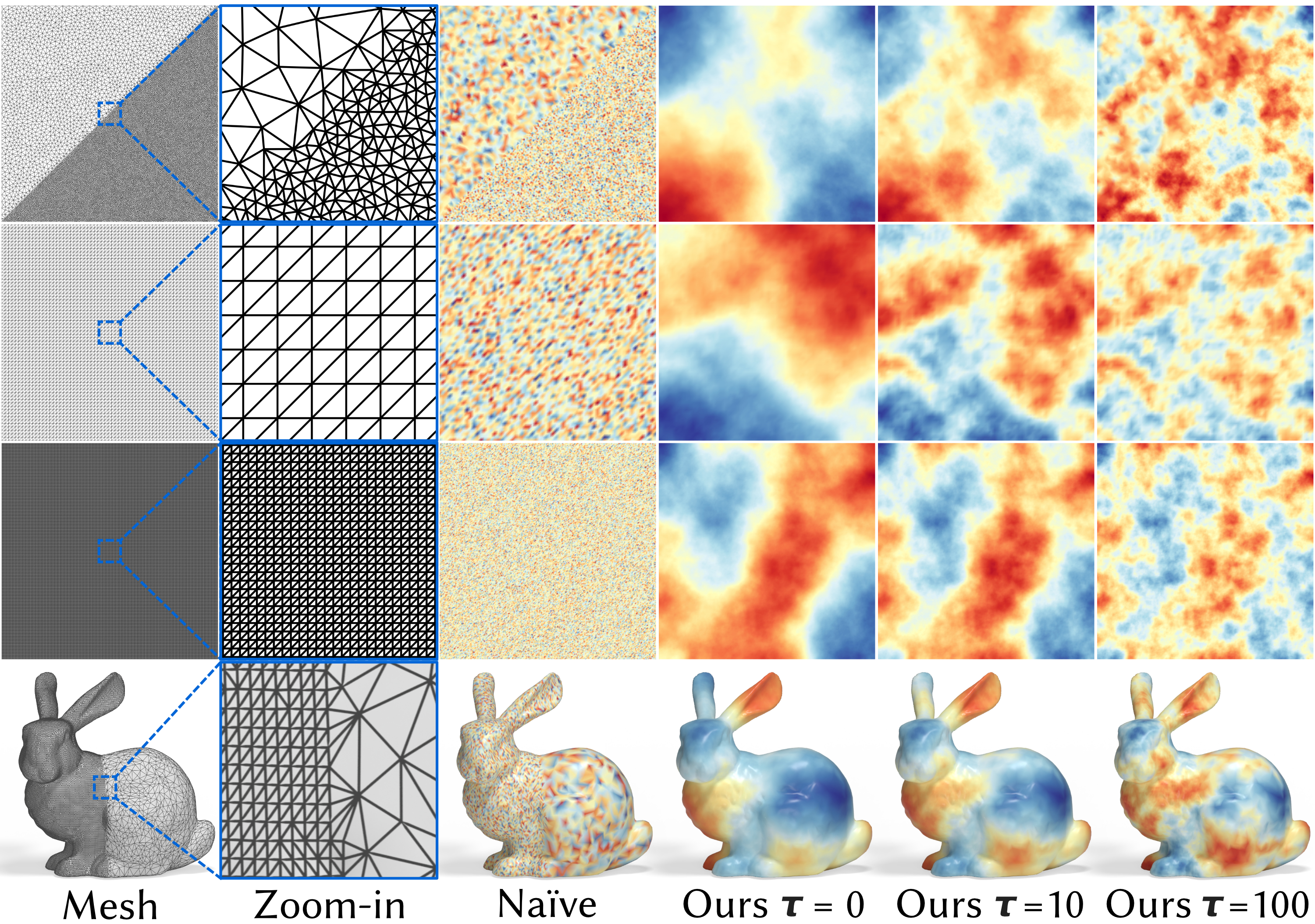}
    \caption{\textbf{\matern{} noise on different triangulations.} 
    {Our} noise exhibits similar structure across differently triangulated regions, as opposed to \textit{Naïve} sampling from iid Gaussians for each vertex.
    \label{fig:noise_on_2d_meshes}
    }
\end{figure}

In designing a triangulation-agnostic \emph{generative framework}, we observe that simply using an existing triangulation-agnostic network architecture is insufficient to obtain agnosticism at sampling time. The primary insight of this paper is that one requires an appropriate \emph{noising scheme} that, like said architectures, is not affected by the specific discretization of the surface. We show that the conventional method of  sampling iid Gaussian noise leads to generative models that cannot operate on unseen triangulations of their training data, let alone generalize to out-of-distribution shapes.

This problem setting leads to the main technical challenge of this paper: while devising a noise distribution, e.g., for images, is a trivial task, the same cannot be said for meshes, considering the aforementioned intricacies; see for example bottom of Figure~\ref{fig:noise_on_2d_meshes}. In this setting, the distribution itself needs to be agnostic to triangulation, lest its characteristics change under superficial changes to the mesh, resulting in noise samples that are out-of-distribution for the trained denoiser (see Figure~\ref{fig:ablations}).

We formalize a definition of triangulation-agnostic noise distributions through the lens of probabilistic spectral shape analysis, allowing us to compare distributions defined on different triangulations. We then observe that a specific type of \emph{\matern{} process}~\cite{matern1960spatial,whittle1963stochastic} has a well-studied linear finite-elements discretization~\cite{lindgren2011explicit} that exactly describes distributions with the desired properties, and leads to a highly simple and efficient sampling scheme using tools that are customary in geometry processing. In particular, one samples from the discrete \matern{} process by simply solving a \textit{screened Poisson equation}, see Algorithm~\ref{alg:sampling}.

\begin{figure}[t]         
    \centering      \includegraphics{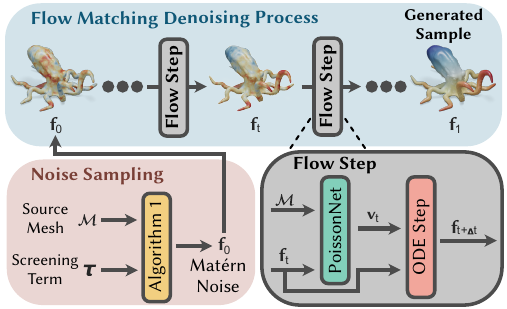}
    \caption{\textbf{Architecture of our triangulation-agnostic generative pipeline.} We first perform \textit{noise sampling} (bottom left) from our proposed triangulation-agnostic \matern{} noise distribution via Algorithm~\ref{alg:sampling}. We then use that noise as the initial signal $\signal_0$ in a \textit{flow matching denoising process} (top), which iteratively denoises the signal into the final sample $\signal_1$. In each \textit{flow step} (bottom right) we input the current signal $\signal_\text{t}$ into  PoissonNet~\cite{maesumi2025poissonnet}, which predicts the current momentary ``velocity'' $\textbf{v}_{\text{t}}$. An ODE step time-integrates the velocity, starting from the current signal $\signal_\text{t}$, to obtain the next signal $\signal_{\text{t}+\Delta \text{t}}$. This continues until reaching $\text{t}=1$.}
    \label{fig:pipeline}
\end{figure}

With this sampling method at hand, we produce a full  triangulation-agnostic generative pipeline based on the flow matching~\cite{lipman2022flow} paradigm.
As a denoising network, we use PoissonNet~\cite{maesumi2025poissonnet}, an architecture that performs triangulation-agnostic learning in the gradient domain. See Figure~\ref{fig:pipeline} for our complete pipeline.

We demonstrate the benefits of our framework on a set of canonical shape editing tasks that are widely studied. In particular, we use our method to generate plausible deformations of humanoid characters as well as resting states of elastic objects in equilibrium. In all cases, our method produces plausible results even on meshes with over one million vertices. By comparison, prior methods in this space are unable to generalize beyond their training meshes (due to sensitivity to triangulation), and in some cases are too inefficient for generation on high-resolution geometry.

To summarize, the main contributions of this work are:
\begin{enumerate}[leftmargin=0.5cm]
    \item The first generative model over meshes that is triangulation-agnostic and applicable to high-resolution meshes.
    \item A theoretical treatment and definition of triangulation-agnostic distributions of signals over meshes via probabilistic analysis of the spectrum of the signals.
    \item Introduction of Matérn processes~\cite{matern1960spatial,whittle1963stochastic} and their discretization~\cite{lindgren2011explicit} as an efficient algorithm for sampling from triangulation-agnostic distributions.
\end{enumerate}
Our code, model checkpoints, datasets and data-generation scripts are available at \href{https://github.com/kts707/matern-fm}{\color{blue}https://github.com/kts707/matern-fm}
\section{Related Work}

\mypar{Generation through denoising.}
Our method builds on \emph{denoising-based} generative modeling, where one starts from a simple noise distribution and iteratively transforms it into an empirical data distribution. We adopt \emph{flow matching} (FM)~\cite{lipman2022flow}, which has been applied to many modalities~\cite{li2025flow,Liuflowaudio}. FM considers a time-varying signal $\signal_t$ parametrized by $t\in\brac{0,1}$, which interpolates between noise $(t=0)$ and data $(t=1)$. Practically, FM predicts a time-dependent vector field, parametrized by a neural network $F_\theta(\signal_t, t)$, which aims to predict the velocity of the signal w.r.t time. Then, at inference time, one generates a sample by integrating the corresponding ODE:
\begin{equation}
    \label{eq:fm}
    \frac{\partial}{\partial t}\signal_t = F_\theta(\signal_t, t),
\end{equation}
from $t=0$ to $t=1$. FM is closely related to \emph{diffusion/score-based} models~\cite{ho2020denoising,song2020score}, which learn to reverse a noising process, as well as ODE-based methods such as rectified flows~\cite{liu2022flow}. We focus on FM and contribute a triangulation-agnostic noising and denoising scheme for signals on meshes.

\mypar{Generation over existing meshes.}
Our work focuses on generating signals \textit{on existing meshes, without modifying their underlying structure}. To the best of our knowledge, there are only two prior works that use current state-of-the-art denoising-based generative techniques~\cite{wang2025doublediffusion,elhag2023manifold}; however, they are not triangulation agnostic. We compare to these methods and show that our method exceeds them in performance.  Slightly older methods use variational autoencoders~\cite{tan2018variational,tan2021variational,yuan2020mesh}; however, this architecture currently does not work sufficiently well for triangulation-agnostic generation. Dupont et al.~\shortcite{dupont22} focus on learning distributions of functions, via a hypernetwork, and can be applied to meshes.

Deformations are a common application of learning-based methods on meshes~\cite{aigerman2022neural,maesumi2023explorable,maesumi2025poissonnet,liu2021deepmetahandles,muralikrishnan2024temporal, sundararaman2024deformation, besnier2024pandas}. They can be generated using VAEs~\cite{muralikrishnan2022glass,huang2021arapreg}, or by optimizing an objective defined by pre-trained vision transformers~\cite{michel2022text2mesh,gao2023textdeformer,kim2025meshup,dinh2025geometry,wang2025headevolver}. Generative techniques can be applied to  skeletal rigs, as opposed to the mesh itself~\cite{gat2025anytop,tevet2023human,shafir2024human,zhong2025sketch2anim}.  We note our approach can be considered a generative method on graphs~\cite{guo2022systematic,zhang2023survey}, though its machinery is heavily tailored for meshes.
\mypar{Non-Euclidean generative modeling.} 
Many works compute flows \textit{over} a manifold in order to sample a \textit{point} on it~\cite{mathieu2020riemannian,chen2023flow,de2022riemannian,huang2022riemannian}. This is  a significantly different setting than ours: we define flows in the \textit{functional space} of the manifold, in order to sample a  \textit{function} defined on the manifold.

\mypar{Generative methods for other 3D modalities.}
The most common generative task for other 3D modalities is generation from scratch (e.g., from a given text prompt). Methods for this task have been proposed for most shape representations: tetrahedra~\cite{gao2022get3d,liu2023meshdiffusion,shen2021dmtet},  voxels~\cite{ren2024xcube,xiang2025structured}, point clouds~\cite{yang2019pointflow,vahdat2022lion,ren2024tiger}, 3D Gaussian Splats~\cite{tang2024lgm}, Neural Radiance Fields~\cite{poole2022dreamfusion}, and hybrid approaches~\cite{zhang2024clay,yan2025object}. 
Generation of meshes~\cite{gao2019sdm,nash2020polygen} can be achieved via LLM agents~\cite{lu2025ll3m} or GPT architectures~\cite{siddiqui2023meshgpt}. Importantly, these works synthesize \emph{shape representations} (geometry and topology) from scratch or from textual/visual prompts. Our problem setting assumes an \emph{existing mesh} and samples from distributions of signals defined on it.

\mypar{Spectral mesh analysis.} We use spectral mesh analysis via eigenvectors of the mesh Laplacian. This is a cornerstone tool in geometry processing~\cite{vallet2008spectral,zhang2010spectral,levy2010spectral,karni2000spectral, ben2005optimality, levy2006laplace}, with various applications~\cite{lescoat2020spectral,mejia2017spectral,song2014mesh}, notably parameterization~\cite{gotsman2003fundamentals}, shape signatures~\cite{sun2009concise},  matching~\cite{ovsjanikov2012functional,zhuravlev2025denoising, pierson2025diffumatch, rimon2025fridu}, analysis~\cite{liu2017dirac,fumero2020nonlinear}, and learning~\cite{smirnov2021hodgenet,sharp2022diffusionnet}. We use spectral analysis in a probabilistic way to define triangulation-agnostic distributions.

\mypar{\matern{} processes.}
Our noise distribution is part of a class of Gaussian random fields called \matern{} processes~\cite{matern1960spatial, whittle1963stochastic}, which are a commonly used statistical tool~\cite{stein1999interpolation,bolin2023statistical} applied in various statistical modeling tasks~\cite{gillan2025discrete,coveney2020gaussian,maddix2021modeling}, as well as in machine learning~\cite{williams2006gaussian}. We use its discretization via linear finite elements, which has been heavily studied~\cite{lindgren2011explicit} (see~\cite{borovitskiy2021matern,borovitskiy2020matern,rosa2023posterior,bolin2024gaussian} for other discretizations).   
All of the above focus on elaborate statistical modeling; we use a specific \matern{} process in a simple way: we observe and prove that it is an ideal model for triangulation-agnostic noise.

\mypar{Non-trivial noise models.} While iid Gaussian sampling is the default noise distribution for generative techniques,  several works~\cite{nachmani2021non, jolicoeur2023diffusion, huang2024blue} have explored alternative initial noise distributions on toy or image datasets. In graphics, noise is commonly used in procedural generation~\cite{perlin1985image}, and point sampling~\cite{ahmed2022gaussian}.
\section{Triangulation-Agnostic Distributions}
\label{ss:noise}

\mypar{Overview.} In this section we define triangulation-agnostic distributions (Section~\ref{ss:properties}) using a spectral perspective, then we identify that a ~\matern{} process  matches that definition in the continuous case (Section~\ref{ss:matern}), and finally we describe how to compute samples from its FEM discretization and review its properties (Section~\ref{ss:sample}).  In Section~\ref{s:pipeline}, we then utilize this distribution as the noise scheme in a  flow matching~\cite{lipman2022flow} pipeline using a triangulation-agnostic denoising network---see Figure~\ref{fig:pipeline}.

\subsection{Preliminaries}

Let $\mesh$ be a manifold triangle mesh discretizing a smooth 2-manifold $\manifold$
with lumped mass matrix $\mass$ and cotangent Laplacian $\lap$~\cite{pinkall1993computing} (see the supplemental material, Section~\ref{s:fem} for an elaboration).  A \emph{signal} is a scalar function sampled at vertices, represented by $\signal\in\mathbb{R}^{n}$. 
We use the mass-weighted inner product between two signals,
$\tribrac{\signal,\textbf{g}}_\mass=\signal^\top \mass\textbf{g}.$
The Laplacian $\lap$ yields generalized eigenpairs, $(\evec,\lambda)_i$, satisfying 
\begin{equation}
\label{eq:geneig}
    \mass^{-1}\lap\evec_i = \lambda_i\evec_i,
\end{equation} with $\mass$-orthonormality $\tribrac{\evec_i,\evec_j}_\mass = \delta_{ij}$. Let $\evecmat$ be the matrix with the eigenvectors as columns. 

We will heavily use the \textit{spectral} representation of a signal $\signal$, which  is its representation as the (unique) linear combination of the Laplacian's eigenvectors, $\signal = \sum_{i=1}^n \coef_i \evec_i$,
where $\coef\in\real^n$ is a vector of coefficients corresponding to each eigenvector.
$\coef_i$ can be directly computed by projecting  $\signal$ onto the eigenvector $\evec_i$:
\begin{equation}
\label{eq:projection}
    \coef_i = \tribrac{\signal,\evec_i}_\mass=\signal^\top \mass \evec_i.
\end{equation} 

The spectral representation is widely used in geometry processing as a generalization of the discrete Fourier transform (DFT) for a 1D signal, with $\evec_i$ analogous to 1D sinusoids, eigenvalues $\lambda_i$ to frequencies, and the coefficients $\coef$  to amplitude of each frequency. The spectrum of the Laplace-Beltrami operator (LBO) $\Delta$ of a compact manifold $\manifold$ is discrete and infinite, and the spectrum of the cotangent Laplacian $\lap$ of a mesh with $n$ vertices is assumed to approximate its $n$ lowest eigenmodes~\cite{wardetzky2008convergence,wardetzky2007discrete}.

We will consider multivariate normal distributions $\normaldist{\zvec}{\covarmat}$, with zero mean and $\covarmat$ covariance. 
 We will make use of the basic fact that if $x\sim \normaldist{\zvec}{\Sigma}$, and $y=Ax$ for some linear transformation $A$, then $y$'s distribution is a normal distribution equal to:
 \begin{equation}
 \label{eq:linear_normal}
     y\sim\normaldist{\zvec}{A^T\Sigma A}.
 \end{equation}

\subsection{Problem Statement}
Formally, we seek an algorithm that, for a given mesh $\mesh$, samples from a distribution of signals defined over its vertices, $\signal\sim\dist^\mesh$.
We refer to such a randomly-sampled signal as a \emph{random field}.
The resulting distribution from which samples are drawn should be triangulation agnostic, meaning that for two triangulations of the same surface, $\mesh\approx \manifold,\ \othermesh\approx\manifold$, the sampling procedure should yield similar distributions $\dist^\mesh\approx\dist^\othermesh$.

\begin{figure}[t]         
    \centering      \includegraphics{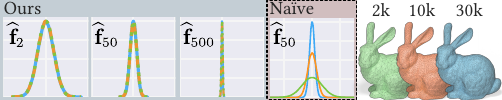}
    \caption{\textbf{Empirical per-frequency statistics of \matern{} noise.} For each of three spectral coefficients -- $\coef_2,\coef_{50},\coef_{500}$ -- we compute its distribution over 100k sampled noises, visualized as a histogram. We show overlays of the distributions for three mesh resolutions. Our method generates near-identical distributions for all meshes, with decreasing variance for higher coefficients. \textit{Naïve}, iid sampling of Gaussians, yields different distributions for each mesh. }
    \label{fig:distributions}
\end{figure}

Since the two meshes have a different number of vertices with no given correspondence between them, a direct comparison between distributions requires notions such as a 1-to-1 map between the meshes and earth mover's distance, which would obstruct our attempt to construct a triangulation-agnostic distribution \textit{a priori}. 

Hence, we compare distributions in the spectral domain, where \emph{frequency} provides a canonical one-dimensional coordinate system. 
Namely, the spectral coefficients $\coef$ are a function of $\signal$, and hence are also random variables distributed according to an (unknown) distribution $\spectraldist^\mesh$.  Intuitively, $\coef$ is a ``dual random field'' over the spectrum, and $\spectraldist^\mesh$ assigns a probability density for any specific spectral coefficients. 
Our goal is to construct a distribution $\dist^\mesh$ s.t. its spectral distribution $\spectraldist^\mesh$ has coefficients that yield consistent statistics across triangulations and stable behavior under mesh refinement (Figure~\ref{fig:distributions}). For simplicity, we will henceforth omit $\mesh$ and refer to the distributions as $\dist,\spectraldist$ when unambiguous.

\subsection{Spectral Properties for Triangulation Agnosticism}
\label{ss:properties}

We  define a triangulation-agnostic distribution $\dist$ as satisfying three key spectral properties in terms of the statistical behavior of the spectral coefficients $\coef$ of signals sampled from  that distribution:

\begin{enumerate}[leftmargin=0.5cm]
    \item \textbf{Per-frequency independence.}
    The coefficients, $\coef_i$, are statistically mutually independent for all $i$, each sampled from an independent univariate distribution.
    \label{item:ind}

    \item \textbf{Mesh-invariant frequency statistics.}
    There exists a family of univariate distributions  $\spectraldist_x$,  parameterized by $x$, such that each spectral coefficient $\coef_i$ with eigenvalue $\lambda_i$ is distributed as $\coef_i\sim\spectraldist_{\lambda_i}$, regardless of the mesh. Moreover, $\spectraldist_x$ is Lipschitz with respect to $x$; specifically, for $|\lambda-\lambda'|\leq \varepsilon$, $$W_2(\spectraldist_\lambda,\spectraldist_{\lambda'}) = \mathcal{O}(\varepsilon),$$
    where $W_2$ is the 2-Wasserstein distance. Note that this should hold between \emph{any two meshes}, even ones that represent different objects or have different genus.
    \label{item:agnostic}

    \item \textbf{Bounded high-frequency content.}
    For any $\varepsilon>0$, there exists a $k$ such that the tail contribution beyond frequency $k$ is negligible:
    $$
    \variance{\sum_{i=k}^{n}\coef_i}<\varepsilon,
    $$
    up to the usual discretization error in approximating the continuous spectrum.
    \label{item:decay}
\end{enumerate}
To give insight into the necessity of these properties:
\begin{itemize}[leftmargin=0.5cm]
    \item Property~\ref{item:ind} decouples behavior across frequencies and enables element-wise comparisons between univariate distributions.

    \item Property~\ref{item:agnostic} enforces that each coefficient's univariate distribution is specified in a mesh-independent manner, purely as a function of frequency, behaving the same across any two arbitrary meshes (see Figure~\ref{fig:distributions}). This makes the distribution stable under re-triangulation and spectral perturbations,  and prevents resolution changes from altering the distribution, except through predictable high-frequency additions controlled by Property~\ref{item:decay}.

    \item Property~\ref{item:decay} prevents higher-resolution meshes from injecting unbounded additional high-frequency energy, as the iid gaussian noise does in Figure~\ref{fig:noise_on_2d_meshes}. See also Figure~\ref{fig:spectrum_illustration} for an illustration of undesired behavior under addition of new vertices (red region) of white noise (unbounded energy) vs. our noise (decaying energy per frequency).
\end{itemize}

We next identify a distribution that satisfies these properties.

\subsection{\matern{} Processes}
\label{ss:matern}
We observe that there exists a specific distribution that satisfies the desired properties from  Section~\ref{ss:properties}. 
In the continuous case, it is defined as a distribution over functions $f: \manifold\to\real$ via the stochastic PDE:
\begin{equation}
\label{eq:matern}
    \Delta f +\tau  f = \mathcal{W},
\end{equation}
where $\tau>0$ is a user-chosen parameter, and $\mathcal{W}$ is a sample from a white noise distribution, i.e., sampling iid from a normal distribution for all points on $\manifold$. 
\newcommand{\insetSpectrum}{
\setlength{\intextsep}{0pt}
\setlength{\columnsep}{0.5em}
 \begin{wrapfigure}{r}{90pt}
   \centering
   \includegraphics[width=\linewidth]{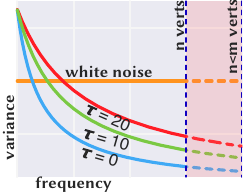}
   \caption{Spectra of the different noise choices.}
   \label{fig:spectrum_illustration}
 \end{wrapfigure}
}

\insetSpectrum In geometry processing terms, $f$ is sampled by first drawing a white noise sample $\mathcal{W}$ (which has iid Gaussian spectrum), and then solving a \emph{screened Poisson equation} with respect to it, which acts as a low-pass filter for the spectrum.

For the sake of our application, the screening term $\tau$  serves to introduce higher frequency content as desired (see Figures~\ref{fig:spectrum_illustration} and~\ref{fig:noise_on_2d_meshes}): as $\tau$ is raised, the relative variance of the distribution of ${\coef}$ increases; 
however, asymptotically, as we go up the spectrum, eventually $\lambda_i$ becomes the dominant term and the variance tends to zero. We use $\tau=100$ in all experiments.

The above distribution is an instance of a class of Gaussian random fields called \emph{\matern{} processes}~\cite{matern1960spatial, whittle1963stochastic}. We will use the straightforward linear FEM discretization of this distribution, which has been extensively studied~\cite{lindgren2011explicit}.

As far as we are aware, \matern{} processes and their discretization are yet to be used for sampling noise for generative tasks, nor introduced into  graphics and geometry processing. Since we need to prove the discretized version of this distribution satisfies the necessary properties for triangulation agnosticism  from Section~\ref{ss:properties}, we next  provide an intuitive review of the construction of the canonical linear FEM discretization of Equation~\eqref{eq:matern} from a geometry processing perspective. By doing so, we hope to also bring more attention in the computer graphics community to this useful tool.

\subsection{Finite-Element \matern{} Noise}
\label{ss:sample}
We next give an intuitive derivation of the FEM discretization of Equation~\eqref{eq:matern} (see Lindgren et al.~\shortcite{lindgren2011explicit} for a much deeper discussion).
Our derivation follows the discussion above: we derive a discrete equivalent of a white noise distribution  $\whitedist$, then we analyze the spectrum of the distribution $\poissondist$ obtained by applying the screened Poisson equation to samples of that white noise.

 For a given mesh $\mesh$, we consider a multivariate normal distribution over its vertices, $\dist\equiv\normaldist{\zvec}{\covarmat}$, with zero mean and $\covarmat$ covariance. The corresponding spectral distribution is also a normal distribution with zero mean, $\spectraldist\equiv\normaldist{\zvec}{\widehat{\covarmat}}$ per Equation~\eqref{eq:linear_normal}, and considering that $\coef$ is a linear transformation of $\signal$, per Equation~\eqref{eq:projection}.

\mypar{Sampling discrete white noise.} We define the white noise distribution as the normal distribution  \begin{equation}
\label{eq:whitedist}
	\whitedist \triangleq \normaldist{\zvec}{\mass^{-1}}, 
\end{equation}
where $\mass$ is the lumped mass matrix. To verify this is indeed also a \emph{spectral} white noise distribution, let $\whitenoise\sim\whitedist$ be a sample, and $\spectralwhitenoise$ its spectral coefficients. From arithmetic of the variance of Gaussians, we get that the spectral covariance matrix is

\begin{equation}
\label{eq:general_var}
\begin{aligned}
 \widehat{\covarmat}_\text{white} &\equiv \variance{\spectralwhitenoise}  \overset{Eq.~\eqref{eq:projection}}{=} 
\variance{\whitenoise^{\top}\mass\evecmat} \overset{Eq.~\eqref{eq:linear_normal}}{=} 
 \evecmat^\top \mass\covarmat \mass \evecmat \\
 &\overset{Eq.~\eqref{eq:whitedist}}{=} \evecmat^\top \mass \mass^{-1} \mass \evecmat =    \evecmat^\top \mass \evecmat \overset{orth.}{=}  \identity.
    \end{aligned}
\end{equation}
Thus we have
\begin{equation}
\label{eq:specwhitedist}
	\spectralwhitedist = \normaldist{\zvec}{\identity}, 
\end{equation}
i.e., the coefficients $\spectralwhitenoise_i$ are sampled iid from a unit Gaussian, leading to a uniform independent behavior across the spectrum, as desired. 

\mypar{Applying the screened Poisson equation.} The linear FEM discretization of the screened Poisson equation from Equation~\eqref{eq:matern} is $\parr{\lap+\tau \mass}\signal = \mass\whitenoise$, where   $\whitenoise\sim\whitedist$ is a white noise sample. The solution to this equation  is given by
\begin{equation}
\label{eq:dpoisson}
    \signal = \parr{\lap+\tau \mass}^{-1}\mass\whitenoise.
\end{equation}
 $\signal$ is thus a random variable, sampled by the above procedure, its distribution being a normal distribution (per Equation~\eqref{eq:linear_normal}, and considering it is a linear function of the white noise) denoted $\poissondist$.

\begin{figure*}[t]
    \centering
    \includegraphics[width=\linewidth, trim=0 0.25cm 0 0]{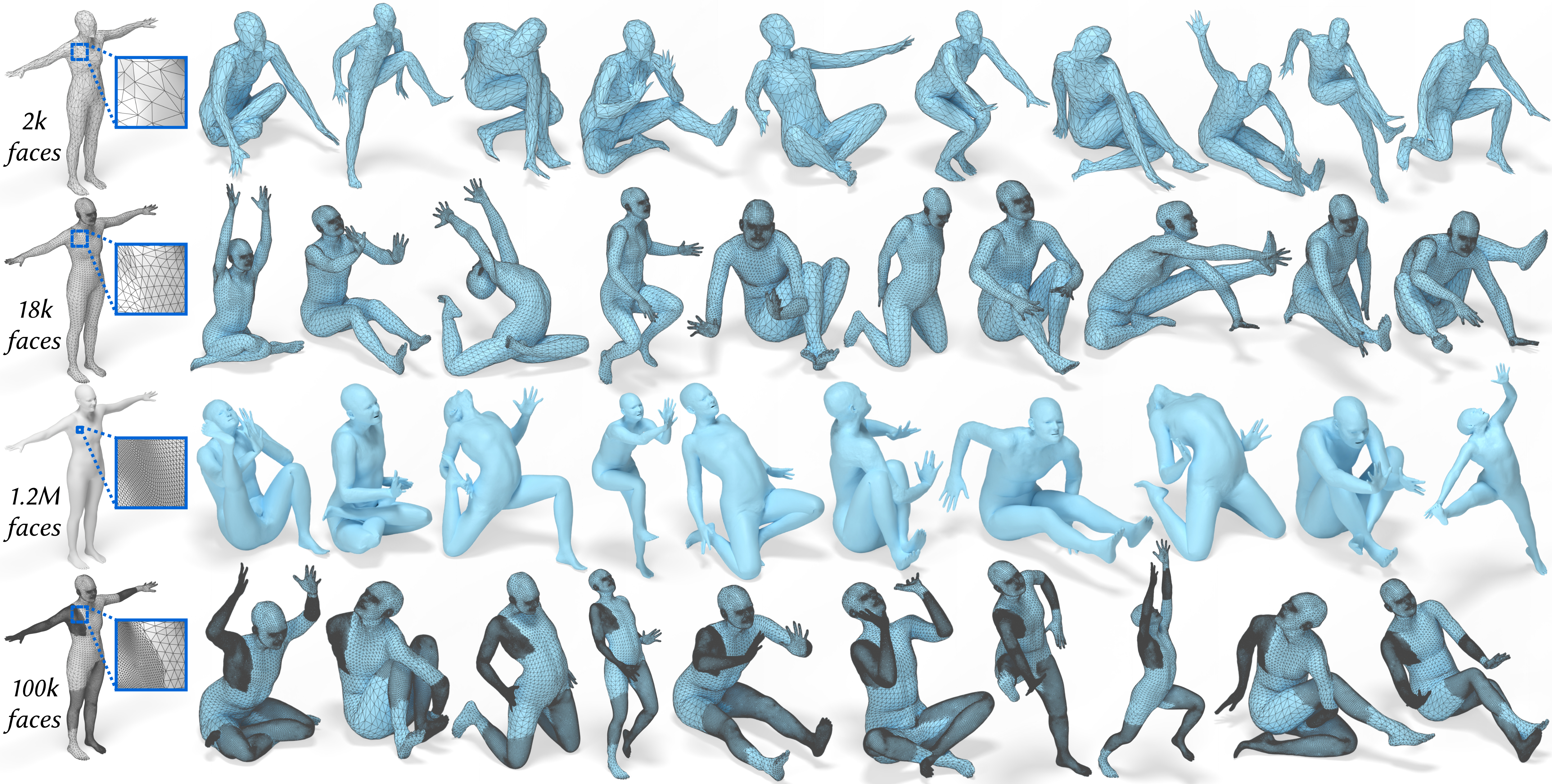}
    \caption{\textbf{Generated reposings}. Our generative model, trained only on the 18k mesh, reposes humans to plausible poses solely via its learned representation.}
    \label{fig:fixed-smpl-src}
\end{figure*}

Next, note that if $\evec_i$ is a generalized eigenvector of $\lap$ with eigenvalue $\lambda_i$, then it is also a generalized  eigenvector of $\parr{\lap+\tau \mass}^{-1}$, with eigenvalue $\parr{\lambda_i + \tau}^{-1}$. Therefore,

\begin{equation}
\begin{aligned}
    \coef_i =\tribrac{\signal,\evec_i}_\mass &=\tribrac{\parr{\lap+\tau \mass}^{-1}\mass \whitenoise, \evec_i}_\mass
    \\=\evec_i^\top\parr{\lap+\tau \mass}^{-1}\mass\whitenoise&= \frac{\evec_i^\top\mass\whitenoise}{\lambda_i + \tau}
    =\frac{\tribrac{\whitenoise,\evec_i}_\mass }{\lambda_i + \tau}\overset{Eq.~\eqref{eq:projection}}{=} \frac{\spectralwhitenoise}{\lambda_i + \tau}.
\end{aligned}
\end{equation}
Per Equation~\eqref{eq:specwhitedist}, $\variance{\spectralwhitenoise_i}=1$, thus we get $\variance{\coef_i}=\variance{\frac{\spectralwhitenoise}{{\lambda_i+\tau}}}=\frac{1}{\parr{\lambda_i+\tau}^2}$, finally yielding the desired spectral distribution
\begin{equation}
\label{eq:poisson_spectrum}
    \spectralpoissondist \equiv\normaldist{\zvec}{\parr{\evalmat + \tau \identity}^{-2}},
\end{equation}
with $\evalmat$ being the diagonal matrix of eigenvalues of the Laplacian. 
In the supplemental material, Appendix~\ref{ap:properties_proof},  we show this spectral distribution satisfies the necessary properties for triangulation agnosticism from Section~\ref{ss:properties}, entailing that $\poissondist$ is a  triangulation-agnostic distribution, as desired.
On a high level: the covariance matrix is diagonal entailing iid sampling, the variance of    each spectral coefficient $\coef_i$ is solely a function of its eigenvalue, and decays fast enough w.r.t. frequency.  

\begin{algorithm}[t]
    \SetKwInOut{Input}{Input}
    \caption{Sampling \matern{} Noise}
    \label{alg:sampling}
    \Input{Mass matrix $\mass$, Laplacian $\lap$, screening term $\tau$. }
    Sample an iid Gaussian random field $\mathbf{n}\sim\normaldist{\zvec}{\identity}$.
    
    Solve $(\lap+\tau \mass)\signal = \sqrt{\mass}\mathbf{n}$ for $\signal$ to obtain the noise.\footnotemark \label{line:algsolve}
\end{algorithm}

We summarize the highly simple and efficient process of sampling \matern{} noise in Algorithm~\ref{alg:sampling}.
As the matrix in line~(\ref{line:algsolve}) is sparse and positive-definite, it can be prefactorized through a Cholesky decomposition for rapid sampling.

\footnotetext{This folds both the white noise and the screened Poisson equation into one equation.}

\begin{figure}[t]         
    \centering      \includegraphics[width=\linewidth]{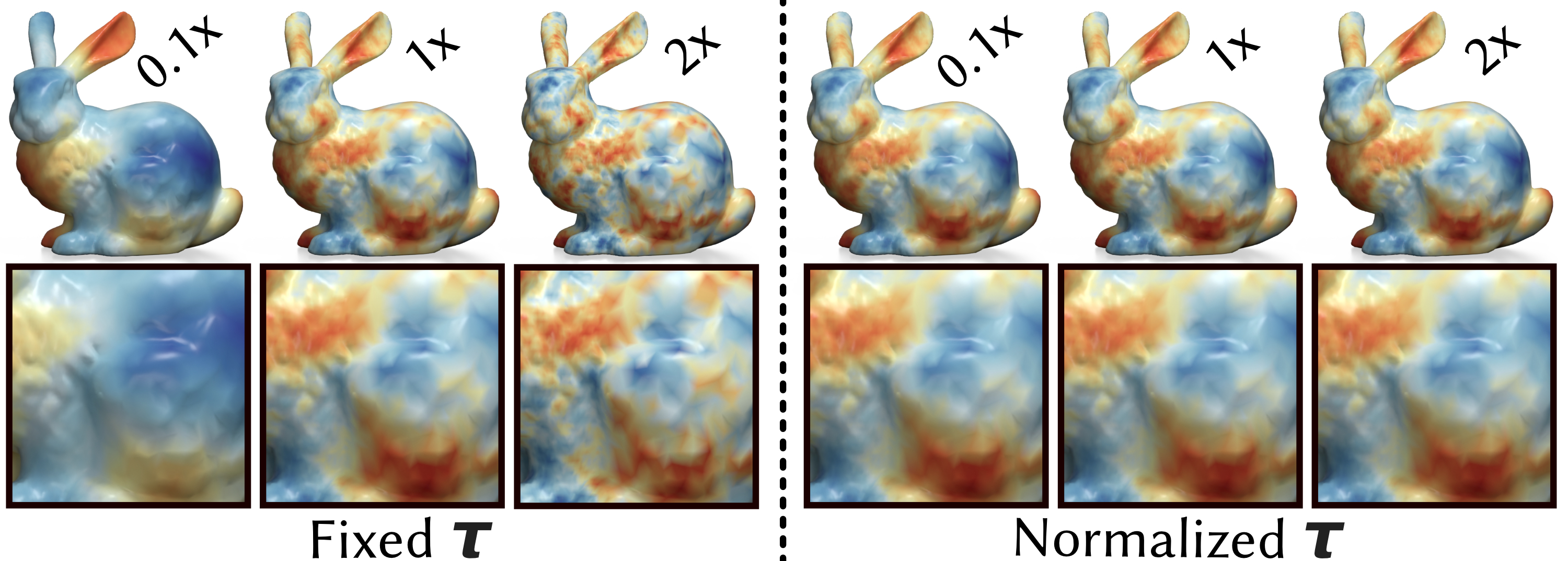}
    \caption{\textbf{Fixed $\tau$ vs. normalized $\tau$.} The \matern{} noise sampled using a fixed screening term $\tau$ is not invariant to the scaling of the mesh (global scaling factors of $0.1$ and $2.0$); with our proposed normalization scheme, the noise exhibits a consistent, scale-independent spectrum.}
    \label{fig:noise_sampling_schemes}
\end{figure}

\mypar{Choosing the screening term $\tau$.} While any specific value of $\tau$ produces consistent noise distributions for different mesh discretizations of the same 3D surface, there is no guarantee of consistent behavior across different shapes, nor between different scaled versions of the same mesh, see Figure~\ref{fig:noise_sampling_schemes}, left. This is due to the change in the spectrum (i.e., eigenvalues and eigenvectors) between different surfaces and between different scales. 

We utilize the Frobenius norm of the discrete Laplace-Beltrami operator, $\Gamma= ||\mass^{-1}\lap||_F$. It is well-known that $\Gamma^2=\sum_i\lambda_i^2$, hence linearly proportional to the eigenvalues. Thus, the user chooses the screening term hyperparameter $c\in\real^+$, and we define $\tau = {c}{\Gamma}$, yielding a screening term $\tau$ proportional to the eigenvalues.
\begin{algorithm}[t]
    \SetKwInOut{Input}{Input}
    \caption{Normalized \matern{} Noise}
    \label{alg:normalize}
    \Input{Mass matrix $\mass$, Laplacian $\lap$, user-chosen  parameter $c$.}
    Compute the normalization factor $\Gamma = ||\mass^{-1} \lap||_F$.
    
    Compute the normalized screening term $\tau = \Gamma c$.
    
    Compute the signal $\signal$ via Algorithm~\ref{alg:sampling}, using $\tau$.

    Return the normalized signal $\sqrt{\Gamma} \signal $.
\end{algorithm}
\begin{figure*}[t]
    \centering
    \includegraphics[width=\linewidth, trim=0 0.25cm 0 0]{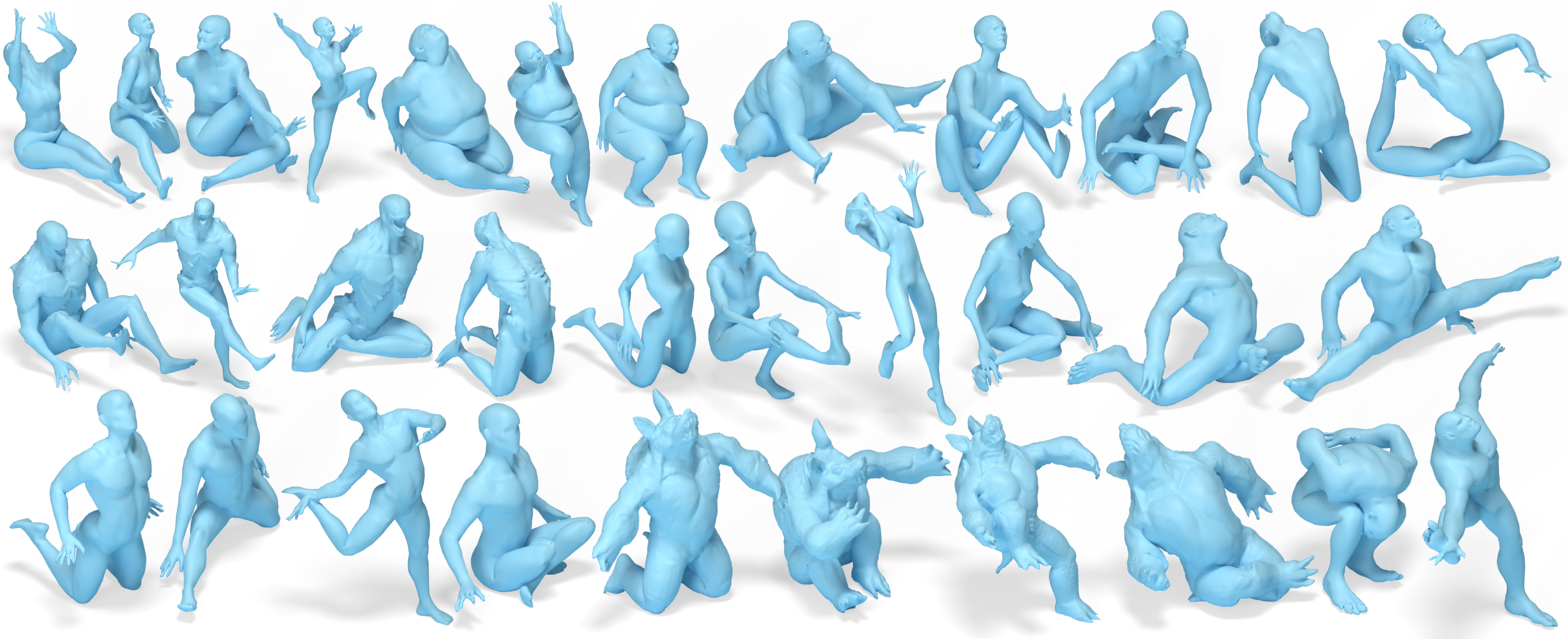}
    \caption{\textbf{Generalization to various 3D shapes}. Our trained model generalizes to various humanoid meshes (never seen during training). }
    \label{fig:multi-smpl-src}
\end{figure*}
\begin{figure}[t]
    \centering
    \includegraphics[width=\linewidth]{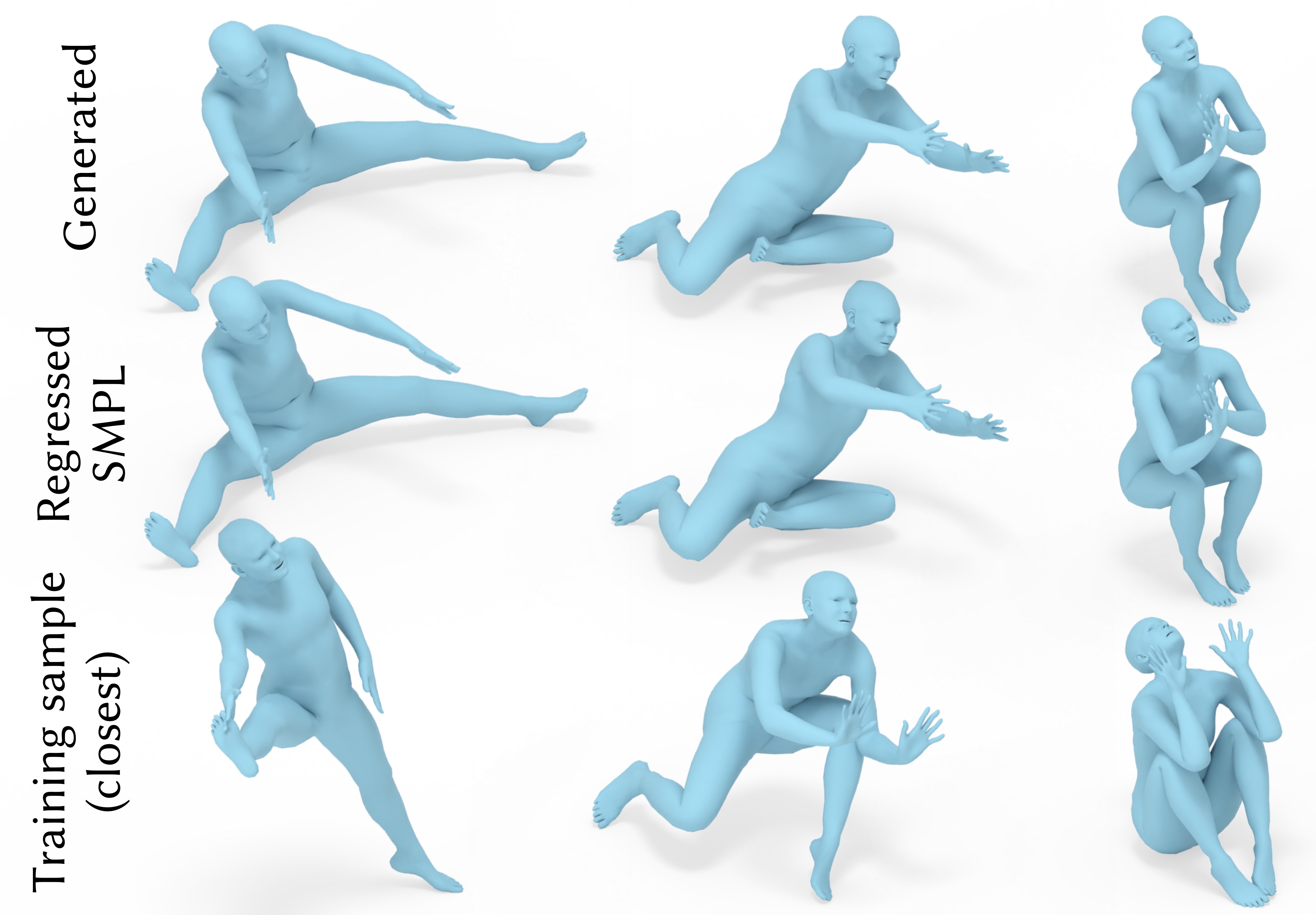}
        \caption{\textbf{Closest training sample and regressed SMPL meshes.}
     Our generated samples approximate the GT deformation distribution well, and can be matched almost perfectly with the SMPL model, via regression. The closest training sample is quite far, showing our method generalizes.
    }
    \label{fig:smpl_closest_training_sample}
\end{figure}

This choice normalizes the screening term w.r.t. the power of the spectrum, leading to more consistent behavior across different meshes, and specifically, to scale invariance; namely , let $\mass^*$,$\lap^*$,$\lambda^*$,$\Gamma^*$, $\signal^*$ be the mass matrix, cotangent Laplacian, eigenvalue, normalizing factor, and signal of the original, unscaled mesh. Assume the mesh's vertex coordinates are scaled by $k\in\real^+$; the resulting (angle-based) cotangent Laplacian remains the same $\lap=\lap^*$; the mass matrix scales up by $k^2$, $\mass=k^2\mass^*$. Hence, per Equation~\eqref{eq:geneig}, $\lambda_i = k^{-2}\lambda_i^*$, and hence also $\Gamma=k^{-2}\Gamma^*$. Plugging this into 
Equation~\eqref{eq:dpoisson} we have 
$\signal = \parr{\lap+ c\Gamma \mass}^{-1}\mass\whitenoise=\parr{\lap^*+ ck^{-2}\Gamma^* k^{2}\mass^*}^{-1}k^{2}\mass^*\whitenoise=\parr{\lap^*+ \tau^* \mass^*}^{-1}k^{2}\mass^*\whitenoise$. 
Per Equation~\eqref{eq:whitedist}, $\whitenoise=k^{-1}\whitenoise^*$, thus we get $\signal = \parr{\lap^*+ \tau^* \mass^*}^{-1}k^{}\mass^*\whitenoise^*=k\signal^*$. Hence, we normalize the resulting signal by multiplying it with $\sqrt{\Gamma} = k^{-1}\sqrt{\Gamma^*}$, thus yielding complete scale invariance: $\sqrt{\Gamma}\signal = \sqrt{\Gamma^*}\signal^*$, see Figure~\ref{fig:noise_sampling_schemes}, right.
We summarize this normalization in Algorithm~\ref{alg:normalize}.

\section{Triangulation-Agnostic Flow Matching}
\label{s:pipeline}

We follow the standard training scheme of rectified flow~\cite{lipman2022flow, liu2022flow, albergo2022building}. Given a mesh with a data sample $\signal_1$ from the dataset, we construct the linear conditional probability paths between a \matern{} noise sample $\signal_0$ and the data sample $\signal_1$ as
        $\signal_t = (1 - t)\cdot\signal_0 + t\cdot\signal_1$,
where $t \in [0,1]$, $\signal_1 \sim \dist_\text{Data}$, and $\signal_0\sim\poissondist$ is sampled from our noise distribution.

To complete the flow matching paradigm, we build our denoising network using a state-of-the-art triangulation agnostic architecture, PoissonNet~\cite{maesumi2025poissonnet}, denoted $F_\theta(\signal_t, t) \to (\textbf{J}_\theta^t, \textbf{v}_\theta^t)$. It operates by first predicting Jacobian velocity $\textbf{J}_\theta^{t}$ in the gradient domain of the signals, which are then translated to vertex velocity $\textbf{v}_\theta^{t}$ via the Poisson equation.
We apply the standard conditional flow matching loss to both vertex and Jacobian velocities as:
\begin{equation}
\mathcal{L}(\theta) = \mathbb{E}_{t, \signal_0, \signal_1} \brac{|| \mass \bigl (\boldsymbol{\delta} - \textbf{v}_\theta^t \bigr)||^2 +  ||\boldsymbol{A} \bigl(\nabla \boldsymbol{\delta} - \textbf{J}_\theta^t \bigr)||^2},
\end{equation}
where $\mass$ and $\boldsymbol{A}$ are the diagonal matrices of vertex masses and face areas, respectively, $\boldsymbol{\delta}=\signal_1 - \signal_0$ is the GT velocity, and $\nabla$ is the \emph{spatial} gradient operator. During the sampling process, we use the vertex velocity $\textbf{v}_\theta^t$ for numerical integration of the ODE (Equation~\eqref{eq:fm}) via the midpoint ODE solver.

\section{Experiments}

\begin{figure*}[t]
    \centering
    \includegraphics[width=\linewidth, trim=0 0.25cm 0 0]{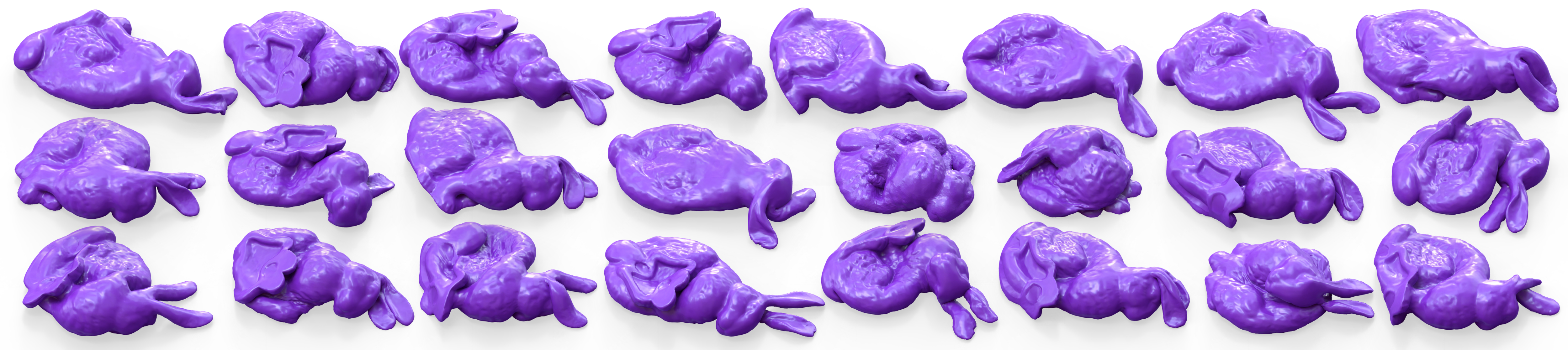}
    \caption{\textbf{Elastic equilibrium states of a deflated bunny}. Our method generates diverse samples on the 70k resolution, though trained solely on 9k resolution. }
    \label{fig:bunny}
\end{figure*}
\begin{figure*}[t]
    \centering
    \includegraphics[width=\linewidth, trim=0 0.25cm 0 0]{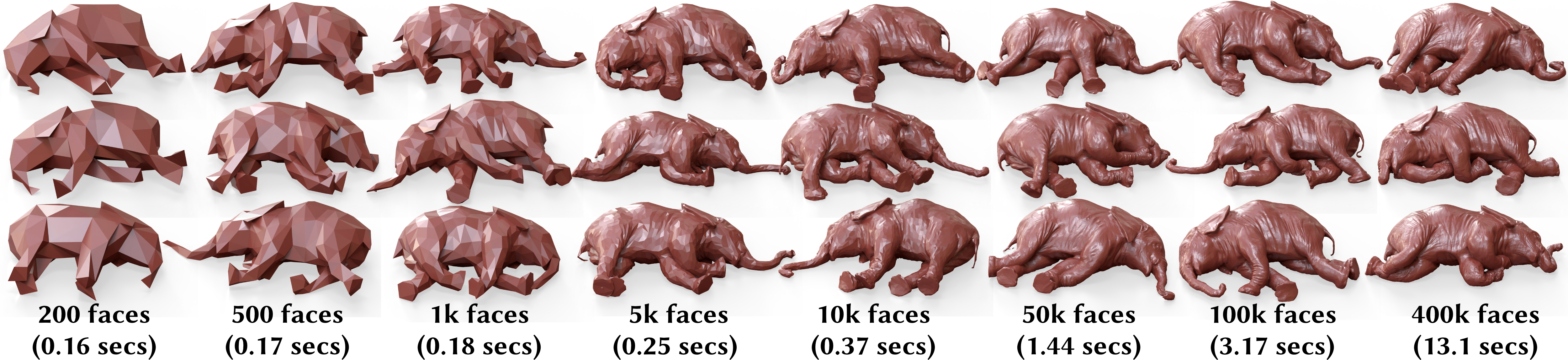}
    \caption{\textbf{Timings for generating elastic rest states over varying resolutions.} Our method produces comparable results on all resolutions, even though it was trained solely on the 10k resolution. Generation time reported underneath each resolution.}
    \label{fig:elephant}
\end{figure*}
\begin{figure*}[t]
    \centering
    \includegraphics[width=\linewidth, trim=0 0.25cm 0 0]{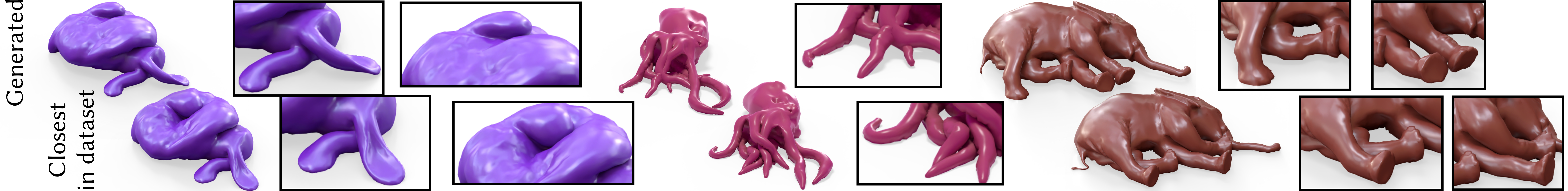}
    \caption{\textbf{Closest sample in the training set to each generated result in the elastic deformation experiment.} }
    \label{fig:phys_closest}
\end{figure*}
\begin{figure}[t]
    \centering
     \includegraphics[width=\linewidth]{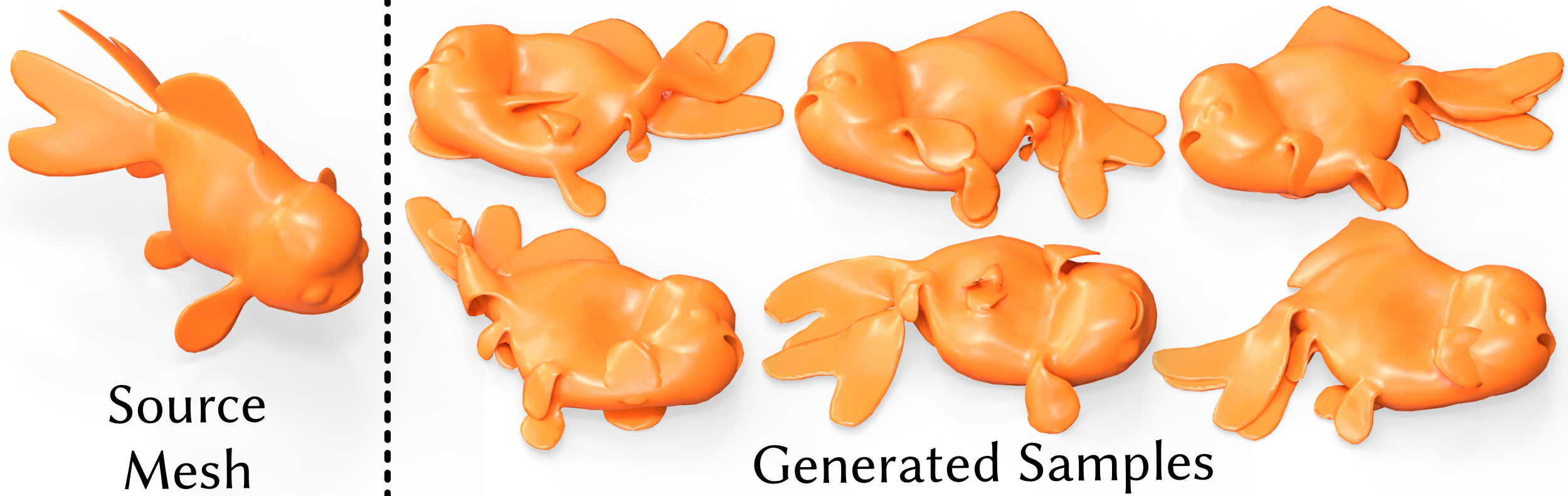}
        \caption{\textbf{Elastic equilibrium of a fish with thin features.} Our method produces plausible deformations even in the thin regions of the fins.}
    \label{fig:fish_results}
\end{figure}

We next experimentally evaluate the capabilities of our generative framework.
In all experiments, we train our model on \textit{one single triangulation}, and validate equivalent performance on various other triangulations, namely of higher-resolution, for which direct training would be infeasible.
\mypar{Deformation datasets.} We choose to focus our experiments on generating deformations, i.e., generating new positions for all vertices of a mesh. This is a central task in learning over meshes~\cite{aigerman2022neural,maesumi2023explorable,muralikrishnan2022glass,maesumi2025poissonnet,gao2023textdeformer,michel2022text2mesh} that fits well to our generative framework, as  datasets are easy to obtain, and our backbone network, PoissonNet~\cite{maesumi2025poissonnet}, has been shown to perform well over it (although thus far, only in a non-generative setting).

\mypar{Evaluation Metrics}
Aside from an error metric which we devise on a per-experiment basis, we additionally estimate how well generated samples approximate the distribution of the dataset via Minimum Matching Distance (MMD) -- the distance to the closest sample in a GT reference set -- and Coverage (COV) -- percentage of samples in reference set that are the nearest to a generated sample; see~\cite{achlioptas2018learning}.

\subsection{Generating Deformations of Humans} 

To evaluate our method against a deformation space with a ground truth parameterization, 
we create a dataset of SMPL~\cite{SMPL:2015} humans in different yoga poses from the MOYO dataset~\cite{ tripathi20233d}, all with a resolution of \emph{18k} faces. Previous works conducted similar reposing experiments~\cite{aigerman2022neural,maesumi2025poissonnet} by training networks w.r.t. SMPL pose parameters;  our method generates deformations without having any awareness of such parameters, making it a much harder task.

\subsubsection{Single source generation.}
\label{sss:singleSMPL}
We train our model to generate deformations of an SMPL human mesh into various yoga poses. In Figure~\ref{fig:fixed-smpl-src}, we show generated samples on different triangulations (up to 1.2M triangles) of the source mesh. Our method consistently generates a diverse set of plausible, elaborate poses.

We check whether the generated deformation is within the SMPL deformation space by regressing the best-fit SMPL parameters to the generated mesh.
More precisely, we deform the SMPL parametric model by simple gradient descent on the SMPL parameters w.r.t. vertex distance to the generated sample. We visualize the regressed SMPL model in Figure~\ref{fig:smpl_closest_training_sample}, and quantitatively compare the generated sample to it via Mean Squared Error (MSE) over the vertices, in Table~\ref{tab:smpl}. Our method produces results that well-conform to the parametric deformation space. Figure~\ref{fig:smpl_closest_training_sample} further shows that the closest sample from the training set to our generated sample is quite far, entailing generalization.

\begin{figure*}[t]
    \centering
    \includegraphics[width=\linewidth, trim=0 0.25cm 0 0]{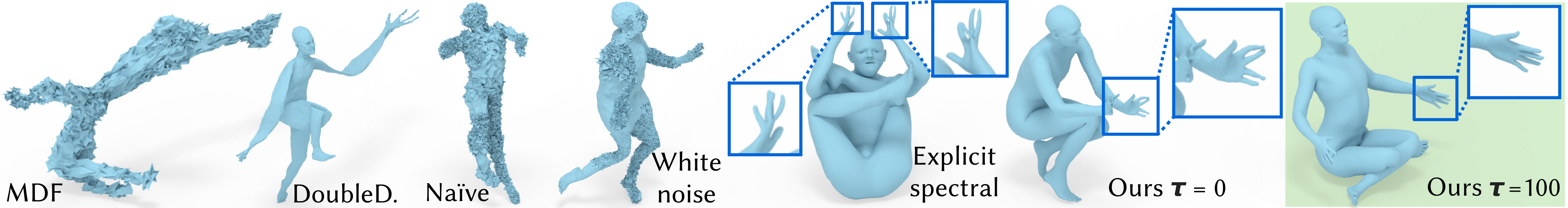}
    \caption{\textbf{Comparisons and ablations}.  We use the reposing experiment (Section~\ref{sss:singleSMPL}) to compare to other mesh-based generative methods: MDF~\cite{elhag2023manifold}, DoubleDiffusion~\cite{wang2025doublediffusion}, the explicit method to compute \matern{} noise, and ablations on other types of noise and screening terms.}
    \label{fig:ablations}
\end{figure*}

\begin{figure}[t]
    \centering
     \includegraphics[width=\linewidth]{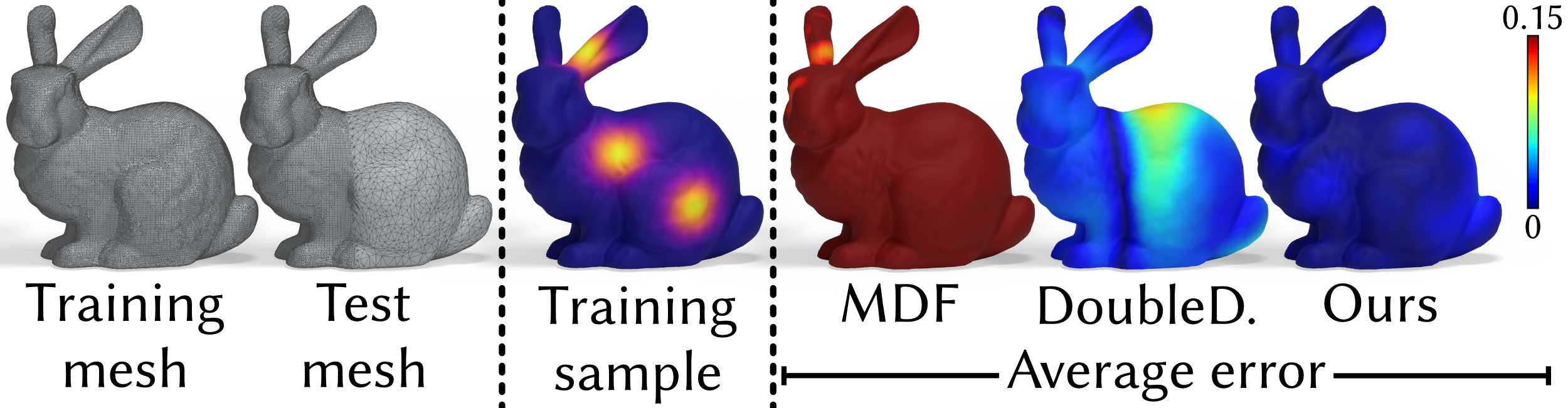}
        \caption{\textbf{Comparisons to MDF~\cite{elhag2023manifold} and DoubleDiffusion~\cite{wang2025doublediffusion} on GMM.} As the competing methods were trained on a regular mesh, MDF and DoubleDiffusion fail to predict correct signals on the irregular test mesh,  as shown by the average error between the GT distribution and the generated samples.}
    \label{fig:gmm}
\end{figure}

\subsubsection{Generation for arbitrary, unseen 3D human shapes.} We use a similar setting as above to produce a model trained to generate deformations for \textit{arbitrary} meshes, representing \emph{different} 3D shapes unseen during training. We create a dataset of  random (source, target) pairs of humans and train the model to generate the target vertices as a signal over the source mesh. To enable different source geometries, we use another PoissonNet~\cite{maesumi2025poissonnet} to predict a positional encoding over the source mesh, which is input as extra channels in the input to the generative model.

As Figure~\ref{fig:multi-smpl-src} shows, this single, trained model can be applied successfully to a wide range of human-like meshes of various shapes, attaining similar diversity and quality to that of the model from Section~\ref{sss:singleSMPL}.

\subsection{Emulating Elastic Deformations}
We evaluate our method on a dataset of  elastic deformations of an object. We generate training data by first creating a volumetric tetrahedral version of the object, and then running IPC~\cite{li2020incremental} to simulate dropping the object at a random initial orientation and angular speed on a ground plane, until it reaches equilibrium.  We then extract the boundary triangle mesh from the deformed tetrahedral meshes and train our triangulation-agnostic flow matching framework on the rest states. Figure~\ref{fig:teaser} shows a set of results on the octopus model, attaining elaborate configurations of its tentacles. 

Note that triangulation agnosticism is critical for this task: when preparing the training set, the tetrahedral mesh used by IPC cannot accommodate the original, high-resolution boundary surface mesh (e.g., 400k faces for the elephant in Figure~\ref{fig:elephant}), hence the training mesh must be of much lower resolution (10k faces).

 Figure~\ref{fig:elephant} further explores the triangulation-agnosticism of our method by generating deformations on a wide range of resolutions.
 We produce plausible results for both extremes of the resolution range (up to the original mesh's resolution). We also report the generation time for each resolution, which is fast enough for real-world applications. In Figure~\ref{fig:bunny}, we show diverse generated samples of a hollow Stanford bunny at its original resolution (70k faces) using a model trained on much lower resolution data (9k faces). See also Figure~\ref{fig:teaser} and Figure~\ref{fig:fish_results}. In Figure~\ref{fig:phys_closest}, we show the closest sample in the training set to a generated sample. For each shape, our model learns and generalizes with plausible variations.

\begin{figure}[t]
    \centering
    \includegraphics[width=\linewidth]{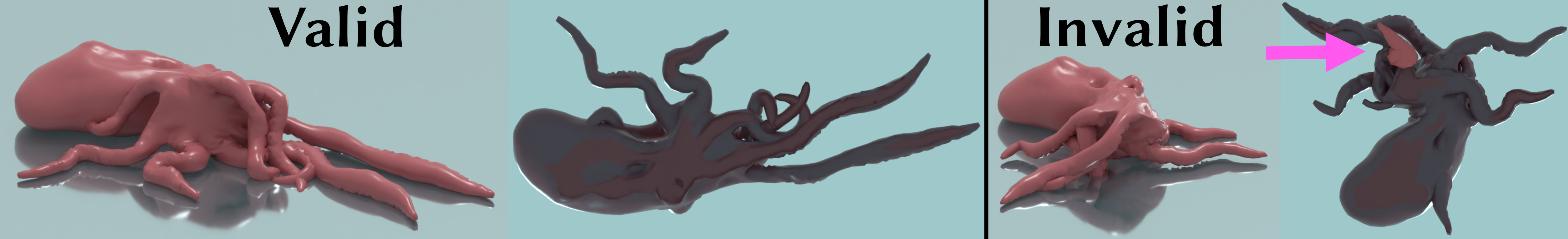}
        \caption{\textbf{Physically inaccurate samples.} Our model generates plausible deformations for the elastic simulation experiment, however can produce  inaccuracies that are physically impossible, e.g., the penetration into the ground plane on the right (seen from underneath), as well as self-penetration.
    \label{fig:failure_case}  
    }
\end{figure}

 We note that we do not aim to perform neural physical simulation~\cite{daviet2025neurally}, as our generated results have small inaccuracies which sometimes lead to collisions and penetration (see Figure~\ref{fig:failure_case}).
 We solely use this dataset due to its very complex and challenging nature for a generative technique.

\begin{table}[t]
\centering
\caption{Quantitative Results on the MOYO~\cite{tripathi20233d} dataset, comparing to DoubleDiffusion~\cite{wang2025doublediffusion} and MDF~\cite{elhag2023manifold},  other noise models, and ablations.}
\label{tab:smpl}
\begin{tabular}{l c c c c}
\toprule
\textbf{Method} & \textbf{MSE}$\,\cdot\, 10^{4}\!\downarrow$ & \textbf{MMD}$\downarrow$ & \textbf{COV}$\uparrow$ & \# \textbf{params}                                                           \\ \midrule

MDF & 1411 & 0.5121 & 0.051 & 11.5M \\
DoubleDiffusion & 155 & 0.1009 & 0.387 & 2.2M \\
\midrule
Naïve  & 2.52 & 0.1365 & 0.210 & 1.4M \\
White noise & 2.91 & 0.1499 & 0.189 & 1.4M \\
Explicit & 0.51 & 0.0744 &0.483 & 1.4M \\
Ours (no screening) & 0.66 & 0.0897 & 0.418 & 1.4M \\
\textbf{Ours (full)}& \textbf{0.27} & \textbf{0.0693} & \textbf{0.487} & 1.4M \\
\bottomrule

\end{tabular}
\end{table}
\begin{table}[]
\centering
\caption{Comparison to MDF~\cite{elhag2023manifold} and DoubleDiffusion~\cite{wang2025doublediffusion} on the synthetic GMM dataset.}
\label{tab:gmm}
\begin{tabular}{l c c c}
\toprule
\textbf{Method} & \textbf{Test Error} $\downarrow$ & \textbf{MMD} ($\times 10^{-2}$) $\downarrow$        & \textbf{COV} $\uparrow$   \\               \midrule
MDF & 0.5245  & 44.67 & 0.012  \\
DoubleDiffusion & 0.0325 & 5.497 & 0.506 \\
\textbf{Ours} & \textbf{0.0071} & \textbf{4.420} & \textbf{0.601} \\
\bottomrule

\end{tabular}
\end{table}
\subsection{Comparisons}

\begin{figure*}[t]
    \centering
    \includegraphics[width=\linewidth, trim=0 0.25cm 0 0]{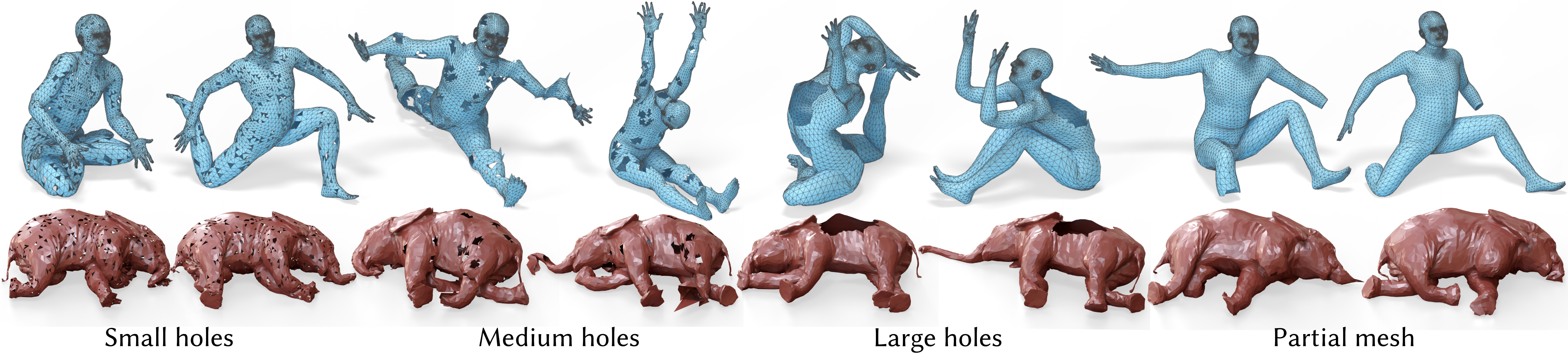}
    \caption{\textbf{Generated results on topologically corrupted source meshes}. Trained solely on the \textit{uncorrupted} source mesh, our method still produces valid results when parts of the source mesh are removed, exhibiting artifacts only near the modified boundaries.}
    \label{fig:corrupted_sources}
\end{figure*}
\begin{figure}[t]
    \centering
     \includegraphics[width=\linewidth]{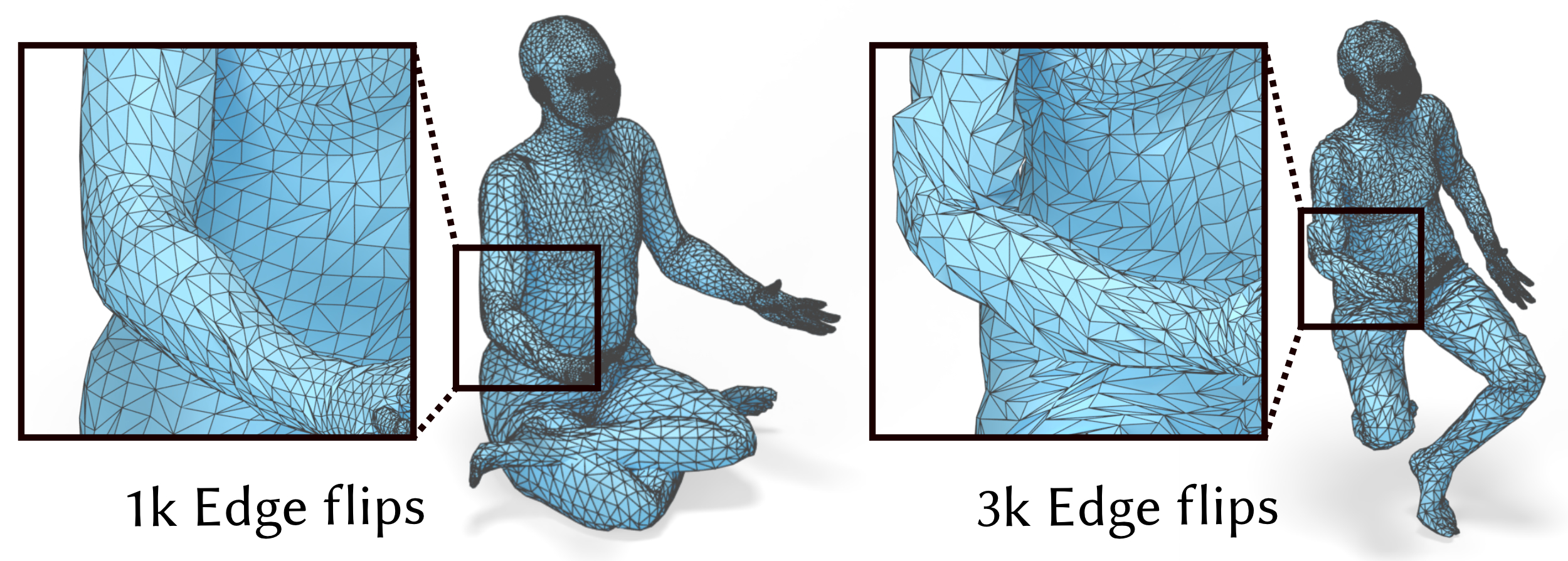}
        \caption{\textbf{Results on source meshes of deteriorated triangulation quality.} We perform 1k and 3k edge flips to create badly triangulated source meshes to test our model, trained solely on \textit{uncorrupted} source meshes.}
    \label{fig:bad_triangles}
\end{figure}

As far as we are aware, there are only two previous works that use modern generative approaches (namely, diffusion models) over meshes: Manifold Diffusion Fields (MDF)~\cite{elhag2023manifold} and DoubleDiffusion~\cite{wang2025doublediffusion}. MDF leverages a transformer-based denoising diffusion model~\cite{ho2020denoising} that operates on inputs positionally encoded via eigenfunctions. DoubleDiffusion uses DiffusionNet~\cite{sharp2022diffusionnet} as the network backbone and trains a diffusion model using the naive iid Gaussian noise, which our experiments show is not triangulation-agnostic. Being the first works in mesh-based generation, they are not triangulation agnostic due to the inherent problem with naive iid Gaussian noise and to the transformer architecture not being triangulation-agnostic. We next compare to both methods in two experiments. We used DoubleDiffusion's publicly available implementation. MDF's code is not released, hence we used the authors' direct advice to instead use the implementation from~\cite{wang2023swallowing} to run their method.

\subsubsection{Generating deformations of human meshes.}

We train MDF and DoubleDiffusion on the human pose generation task (Section~\ref{sss:singleSMPL}) and evaluate them in the same way as our model.
As they are not triangulation agnostic, these methods are significantly less effective in generating diverse and plausible deformations, as Figure~\ref{fig:ablations} and Table~\ref{tab:smpl} show.
Our method exceeds their capabilities both in matching the dataset (MMD, COV) as well as in matching the deformation space (MSE, see Section~\ref{sss:singleSMPL}).

\subsubsection{Generating Gaussian Mixtures}

\begin{table}[t]
\centering
\caption{Our model's performance on source meshes of different corruption types and levels shown in Figure~\ref{fig:corrupted_sources}, evaluated on the MOYO~\cite{tripathi20233d} dataset following the same \textbf{MSE}$\,\cdot\, 10^{4}\!\downarrow$ metric in Table~\ref{tab:smpl}.}
\label{tab:corruption}
\begin{tabular}{ccccccc}
\toprule
\multicolumn{3}{c}{Holes} & \multirow{2}{*}{\begin{tabular}[c]{@{}c@{}}Partial\\ Mesh\end{tabular}} & \multicolumn{2}{c}{\begin{tabular}[c]{@{}c@{}}Deteriorated\end{tabular}} & \multirow{2}{*}{\begin{tabular}[c]{@{}c@{}}Original\\ Mesh\end{tabular}} \\ \cline{1-3} \cline{5-6}
small  & medium  & large  &                                                                         & mild                             & \multicolumn{1}{c}{severe}                            &                                                                          \\
 \midrule
 0.45 & 1.87 & 1.35 & 0.37 & 0.83 & 9.92 & \textbf{0.27} \\
\bottomrule

\end{tabular}
\end{table}

To further show the difference in triangulation-agnosticism between the methods, we recreate MDF's GMM experiment
that learns to generate Gaussian mixtures, by preparing a dataset of Gaussian mixture signals on a regular triangulation of the Stanford Bunny, on which we train all methods. 

At test time, we generate samples on a non-uniformly triangulated test mesh.
We evaluate how accurate the generated signals are w.r.t. the desired GMM by comparing their statistics: we compare the point-wise mean of a large set of generated signals on the mesh vs. the ground truth point-wise mean (evaluated by taking the mean of the training set) -- see Table~\ref{tab:gmm} (column \emph{Test Error}). We outperform the two other methods. As shown in Figure~\ref{fig:gmm}, our method produces samples with correct statistical parameters. MDF shows a general deviation, while DoubleDiffusion shows clear biases in the generated results, correlating with the density of the triangulation.

We focus our comparisons to other generative methods on meshes, as other modalities solve rather different problems. In Appendix~\ref{s:TIGER} in the supplemental material,  we compare our method to a state-of-the-art point cloud generator~\cite{ren2024tiger}, and show the differences and advantages of using a mesh-based method. 

\subsection{Alternatives}
\label{ss:alternatives}
\newcommand{\insetComputeTimePlot}{
\setlength{\intextsep}{0pt}
\setlength{\columnsep}{0.5em}
 \begin{wrapfigure}[12]{r}{80pt}
   \centering
   \includegraphics[width=\linewidth]{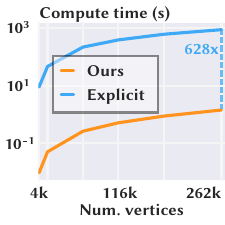}
   \includegraphics[width=\linewidth]{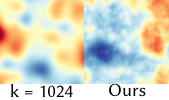}
 \end{wrapfigure}}
\mypar{Explicit spectral decomposition.}  After formally defining triangulation agnostic  distributions (Section~\ref{ss:properties}),  we could have deduced a \insetComputeTimePlot straightforward alternative: \emph{explicitly} compute the first $k$ eigenvectors of the Laplacian and form a random linear combination of them, 
    $\signal = \sum_{i}^{k} \frac{\chi_i}{\lambda_i + \tau} \evec_i,\ \chi_i \sim\normaldist{0}{1}.$
This holds several disadvantages:  
1) The eigendecomposition is computationally expensive (see inset), especially due to the lack of reliable GPU solvers; 2) an eigenbasis of size $1024$ still fails to produce as much high-frequency content as our noise---see inset. This, in turn, leads to less accurate results in Figure~\ref{fig:ablations},  particularly in fine-grained regions, e.g. the fingers. Table~\ref{tab:smpl} verifies this quantitatively. We note that one can use more efficient techniques for computing eigenvectors, such as shift-invert spectral transforms that explore the spectrum band-by-band~\cite{vallet2008spectral}---this may enable other explicit constructions of triangulation-agnostic noise. 

\newcommand{\insetAutoencoder}{
\setlength{\intextsep}{0pt}
\setlength{\columnsep}{0.5em}
 \begin{wrapfigure}[5]{r}{80pt}
   \centering
   \includegraphics[width=\linewidth]{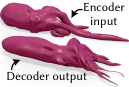}
 \end{wrapfigure}
}
\mypar{Latent generation.} Other generative methods operate in a \emph{latent space}, e.g., variational autoencoders~\cite{KingmaW19}, and, more recently, latent diffusion models~\cite{rombach2022high}. \insetAutoencoder It is unclear whether latent generation outperforms generation in the signal domain~\cite{yu2025pixeldit, li2025back}. Regardless, producing a good latent space in a triangulation-agnostic manner is as of now an unsolved problem, which prevents us from devising a worthy baseline to compare to. To show this,
We trained a \textit{non-variational} autoencoder consisting of a DiffusionNet~\cite{sharp2022diffusionnet} encoder that encodes the input deformation into a $128$ dimensional latent feature and a PoissonNet~\cite{maesumi2025poissonnet} decoder on our octopus deformation dataset. As the inset shows, the quality of the generated result is too poor to be useful.

\subsection{Ablations}

\mypar{Other noise models.} We evaluate the importance of our main contribution, the \matern{} noise, by repeating the reposing experiment from Section~\ref{sss:singleSMPL} using both white noise (Equation~\eqref{eq:whitedist}) and naïve iid Gaussian noise on the vertices. Figure~\ref{fig:ablations} shows the generated deformations over a non-uniform triangulation. The other noise models fail, verifying the necessity of the properties stipulated in Section~\ref{ss:properties}. Table~\ref{tab:smpl} verifies this result quantitatively.

\mypar{Effects of screening term $\tau$.}  Training our model with \matern{} noise without a screening term ($\tau=0$) leads to deterioration in results on the reposing experiment from Section~\ref{sss:singleSMPL}, as shown in Figure~\ref{fig:ablations} and Table~\ref{tab:smpl}. The higher frequency components are essential to the model's success, especially on small parts (e.g. fingers). 

\subsection{Robustness}
We next explore the robustness of the pipeline. Since both the \matern{} noise and the backbone, PoissonNet~\cite{maesumi2025poissonnet} rely on Poisson's equation, the success of our framework hinges on the deterioration in the accuracy of Poisson's equation.
\mypar{Topological corruption.} After training the generative pipeline, we modify the topology of the source mesh by introducing holes, and evaluate the effect on the quality of the results.  As Figure~\ref{fig:corrupted_sources} shows, our method continues to generate plausible results (since Poisson's equation is not severely affected), with some deterioration next to the newly-introduced boundaries. Table~\ref{tab:corruption} shows that the results drift from the SMPL parametric model, as expected. Note that Medium holes lead to higher MSE than Large holes, due to a larger total boundary length of the holes on surface (many medium-sized holes spread over the surface vs. four large-sized holes). 

\mypar{Triangulation quality.} As the triangulation deteriorates in its condition number, both geometry and Poisson's equation are affected. While Figure~\ref{fig:bad_triangles} shows our method still produces plausible results, Table~\ref{tab:corruption} shows that the accuracy of our generated results drops. We note this happens for extremely badly triangulated meshes (see zoom-in in Figure~\ref{fig:bad_triangles}).

\section{Conclusion}
This work takes a step towards practical generative techniques for triangle meshes, and our experiments validate that it produces highly plausible results and is capable of emulating intricate datasets, making it already a viable option for real-world applications.

\mypar{Limitations.} Our method is still not a catch-all generative technique for meshes. A crucial bottleneck lies in the architecture of the denoiser's backbone, PoissonNet~\cite{maesumi2025poissonnet}. As with other triangulation-agnostic architectures~\cite{smirnov2021hodgenet,sharp2022diffusionnet}, it can only tackle tasks that do not require extreme high-frequency content, such as segmentation and deformation, thus preventing us from generating, e.g., crisp RGB images on meshes. Additionally, PoissonNet is limited to operating on single connected-component meshes.  Lastly, the discussed architectures~\cite{sharp2022diffusionnet,maesumi2025poissonnet} are limited to $2$-manifolds (though our noise can be defined in any dimension). Our \matern{} noise can be incorporated into future architectures that eventually emerge to solve these issues.

The \matern{} noise distribution itself holds limitations.
First, it is sensitive to topological changes (See Figure~\ref{fig:corrupted_sources}). Second, we assume nice enough approximations of the mesh~\cite{wardetzky2007discrete,wardetzky2008convergence} for our claims to hold---in practice, we observe that all models used in our experiments produce triangulation-agnostic noise (Figure~\ref{fig:distributions}).

\mypar{Future work.}
In the immediate term,  applications of generative deformations include producing variations of existing 3D models from a given template, retargeting existing poses or physical simulations, and adding details to existing mesh models~\cite{liu2020neural}. 
Extensions of the architecture which seem potentially impactful include generating temporal animations of given meshes and extensions to volumetric data with applications in medical imaging. 

We believe that ~\matern{} processes have much more to offer to the graphics community, both in pure-graphics applications, e.g., modeling  noise~\cite{perlin1985image}, as well as in  non-meshy generative contexts, such as devising pixel-agnostic image generators.

\begin{acks}
This material is based upon work supported by: NSERC Discovery grant RGPIN-2024-04605, “Practical Neural Geometry Processing”; FRQNT Établissement de la relève professorale 365040, “Calcul rapide et léger des déformations à l’aide de réseaux neuronaux”; FRQNT team grant No. 361570; Digital Research Alliance of Canada, Compute Ontario, and Calcul Québec.
\end{acks}

% Bibliography
\bibliographystyle{ACM-Reference-Format}
\bibliography{bibliography}

@String{Computing = "Computing" }

@String{Computer = "{IEEE} Computer" }

@String{Academic = "Academic Press" }

@String{Springer = "Springer-Verlag" }

@article{ho2020denoising,
  title={Denoising diffusion probabilistic models},
  author={Ho, Jonathan and Jain, Ajay and Abbeel, Pieter},
  journal={Advances in neural information processing systems},
  volume={33},
  pages={6840--6851},
  year={2020}
}

@inproceedings{rombach2022high,
  title={High-resolution image synthesis with latent diffusion models},
  author={Rombach, Robin and Blattmann, Andreas and Lorenz, Dominik and Esser, Patrick and Ommer, Bj{\"o}rn},
  booktitle={Proceedings of the IEEE/CVF conference on computer vision and pattern recognition},
  pages={10684--10695},
  year={2022}
}

@article{lipman2022flow,
  title={Flow matching for generative modeling},
  author={Lipman, Yaron and Chen, Ricky TQ and Ben-Hamu, Heli and Nickel, Maximilian and Le, Matt},
  journal={arXiv preprint arXiv:2210.02747},
  year={2022}
}

@article{albergo2022building,
  title={Building normalizing flows with stochastic interpolants},
  author={Albergo, Michael S and Vanden-Eijnden, Eric},
  journal={arXiv preprint arXiv:2209.15571},
  year={2022}
}

@article{chen2023flow,
  title={Flow matching on general geometries},
  author={Chen, Ricky TQ and Lipman, Yaron},
  journal={arXiv preprint arXiv:2302.03660},
  year={2023}
}

@article{mathieu2020riemannian,
  title={Riemannian continuous normalizing flows},
  author={Mathieu, Emile and Nickel, Maximilian},
  journal={Advances in neural information processing systems},
  volume={33},
  pages={2503--2515},
  year={2020}
}

@inproceedings{elhag2023manifold,
  author       = {Ahmed A. A. Elhag and
                  Yuyang Wang and
                  Joshua M. Susskind and
                  Miguel {\'{A}}ngel Bautista},
  title        = {Manifold Diffusion Fields},
  booktitle    = {The Twelfth International Conference on Learning Representations,
                  {ICLR} 2024, Vienna, Austria, May 7-11, 2024},
  publisher    = {OpenReview.net},
  year         = {2024},
  url          = {https://openreview.net/forum?id=BZtEthuXRF},
  bibsource    = {dblp computer science bibliography, https://dblp.org}
}

@article{wang2025doublediffusion,
  title={DoubleDiffusion: Combining Heat Diffusion with Denoising Diffusion for Texture Generation on 3D Meshes},
  author={Wang, Xuyang and Cheng, Ziang and Li, Zhenyu and Yang, Jiayu and Ji, Haorui and Ji, Pan and Harandi, Mehrtash and Hartley, Richard and Li, Hongdong},
  journal={arXiv preprint arXiv:2501.03397},
  year={2025}
}

@article{smirnov2021hodgenet,
  title={HodgeNet: Learning spectral geometry on triangle meshes},
  author={Smirnov, Dmitriy and Solomon, Justin},
  journal={ACM Transactions on Graphics (TOG)},
  volume={40},
  number={4},
  pages={1--11},
  year={2021},
  publisher={ACM New York, NY, USA}
}

@article{sharp2022diffusionnet,
  title={Diffusionnet: Discretization agnostic learning on surfaces},
  author={Sharp, Nicholas and Attaiki, Souhaib and Crane, Keenan and Ovsjanikov, Maks},
  journal={ACM Transactions on Graphics (TOG)},
  volume={41},
  number={3},
  pages={1--16},
  year={2022},
  publisher={ACM New York, NY}
}

@article{maesumi2025poissonnet,
  title={PoissonNet: A Local-Global Approach for Learning on Surfaces},
  author={Maesumi, Arman and Makadia, Tanish and Groueix, Thibault and Kim, Vladimir and Ritchie, Daniel and Aigerman, Noam},
  journal={ACM Transactions on Graphics (TOG)},
  volume={44},
  number={6},
  pages={1--16},
  year={2025},
  publisher={ACM New York, NY, USA}
}

@article{nachmani2021non,
  title={Non gaussian denoising diffusion models},
  author={Nachmani, Eliya and Roman, Robin San and Wolf, Lior},
  journal={arXiv preprint arXiv:2106.07582},
  year={2021}
}

@article{jolicoeur2023diffusion,
  title={Diffusion models with location-scale noise},
  author={Jolicoeur-Martineau, Alexia and Fatras, Kilian and Li, Ke and Kachman, Tal},
  journal={arXiv preprint arXiv:2304.05907},
  year={2023}
}

@article{ahmed2022gaussian,
  title={Gaussian blue noise},
  author={Ahmed, Abdalla GM and Ren, Jing and Wonka, Peter},
  journal={ACM Transactions on Graphics (TOG)},
  volume={41},
  number={6},
  pages={1--15},
  year={2022},
  publisher={ACM New York, NY, USA}
}

@inproceedings{huang2024blue,
  title={Blue noise for diffusion models},
  author={Huang, Xingchang and Salaun, Corentin and Vasconcelos, Cristina and Theobalt, Christian and Oztireli, Cengiz and Singh, Gurprit},
  booktitle={ACM SIGGRAPH 2024 conference papers},
  pages={1--11},
  year={2024}
}

@article{aigerman2022neural,
  title={Neural jacobian fields: learning intrinsic mappings of arbitrary meshes},
  author={Aigerman, Noam and Gupta, Kunal and Kim, Vladimir G and Chaudhuri, Siddhartha and Saito, Jun and Groueix, Thibault},
  journal={ACM Transactions on Graphics (TOG)},
  volume={41},
  number={4},
  pages={1--17},
  year={2022},
  publisher={ACM New York, NY, USA}
}

@inproceedings{maesumi2023explorable,
  title={Explorable mesh deformation subspaces from unstructured 3d generative models},
  author={Maesumi, Arman and Guerrero, Paul and Aigerman, Noam and Kim, Vladimir and Fisher, Matthew and Chaudhuri, Siddhartha and Ritchie, Daniel},
  booktitle={SIGGRAPH Asia 2023 Conference Papers},
  pages={1--11},
  year={2023}
}

@article{yu2025pixeldit,
  title={PixelDiT: Pixel Diffusion Transformers for Image Generation},
  author={Yu, Yongsheng and Xiong, Wei and Nie, Weili and Sheng, Yichen and Liu, Shiqiu and Luo, Jiebo},
  journal={arXiv preprint arXiv:2511.20645},
  year={2025}
}

@article{li2025back,
  title={Back to basics: Let denoising generative models denoise},
  author={Li, Tianhong and He, Kaiming},
  journal={arXiv preprint arXiv:2511.13720},
  year={2025}
}

@article{wang2023swallowing,
  title={Swallowing the bitter pill: Simplified scalable conformer generation},
  author={Wang, Yuyang and Elhag, Ahmed A and Jaitly, Navdeep and Susskind, Joshua M and Bautista, Miguel Angel},
  journal={arXiv preprint arXiv:2311.17932},
  year={2023}
}

@inproceedings{SMPL-X:2019,
  title = {Expressive Body Capture: {3D} Hands, Face, and Body from a Single Image},
  author = {Pavlakos, Georgios and Choutas, Vasileios and Ghorbani, Nima and Bolkart, Timo and Osman, Ahmed A. A. and Tzionas, Dimitrios and Black, Michael J.},
  booktitle = {Proceedings IEEE Conf. on Computer Vision and Pattern Recognition (CVPR)},
  pages     = {10975--10985},
  year = {2019}
}

@article{SMPL:2015,
      author = {Loper, Matthew and Mahmood, Naureen and Romero, Javier and Pons-Moll, Gerard and Black, Michael J.},
      title = {{SMPL}: A Skinned Multi-Person Linear Model},
      journal = {ACM Trans. Graphics (Proc. SIGGRAPH Asia)},
      month = oct,
      number = {6},
      pages = {248:1--248:16},
      publisher = {ACM},
      volume = {34},
      year = {2015}
    }

@inproceedings{tripathi20233d,
  title={3D human pose estimation via intuitive physics},
  author={Tripathi, Shashank and M{\"u}ller, Lea and Huang, Chun-Hao P and Taheri, Omid and Black, Michael J and Tzionas, Dimitrios},
  booktitle={Proceedings of the IEEE/CVF conference on computer vision and pattern recognition},
  pages={4713--4725},
  year={2023}
}

@article{nicolet2021large,
  title={Large steps in inverse rendering of geometry},
  author={Nicolet, Baptiste and Jacobson, Alec and Jakob, Wenzel},
  journal={ACM Transactions on Graphics (TOG)},
  volume={40},
  number={6},
  pages={1--13},
  year={2021},
  publisher={ACM New York, NY, USA}
}

@article{li2020incremental,
  title={Incremental potential contact: intersection-and inversion-free, large-deformation dynamics.},
  author={Li, Minchen and Ferguson, Zachary and Schneider, Teseo and Langlois, Timothy R and Zorin, Denis and Panozzo, Daniele and Jiang, Chenfanfu and Kaufman, Danny M},
  journal={ACM Trans. Graph.},
  volume={39},
  number={4},
  pages={49},
  year={2020}
}

@article{hu2020fast,
  title={Fast tetrahedral meshing in the wild},
  author={Hu, Yixin and Schneider, Teseo and Wang, Bolun and Zorin, Denis and Panozzo, Daniele},
  journal={ACM Transactions on Graphics (ToG)},
  volume={39},
  number={4},
  pages={117--1},
  year={2020},
  publisher={ACM New York, NY, USA}
}

@inproceedings{achlioptas2018learning,
  title={Learning representations and generative models for 3d point clouds},
  author={Achlioptas, Panos and Diamanti, Olga and Mitliagkas, Ioannis and Guibas, Leonidas},
  booktitle={International conference on machine learning},
  pages={40--49},
  year={2018},
  organization={PMLR}
}

@article{lindgren2011explicit,
  title={An explicit link between Gaussian fields and Gaussian Markov random fields: the stochastic partial differential equation approach},
  author={Lindgren, Finn and Rue, H{\aa}vard and Lindstr{\"o}m, Johan},
  journal={Journal of the Royal Statistical Society Series B: Statistical Methodology},
  volume={73},
  number={4},
  pages={423--498},
  year={2011},
  publisher={Oxford University Press}
}

@article{pinkall1993computing,
  title={Computing discrete minimal surfaces and their conjugates},
  author={Pinkall, Ulrich and Polthier, Konrad},
  journal={Experimental mathematics},
  volume={2},
  number={1},
  pages={15--36},
  year={1993},
  publisher={Taylor \& Francis}
}

@book{stein1999interpolation,
  title={Interpolation of spatial data: some theory for kriging},
  author={Stein, Michael L},
  year={1999},
  publisher={Springer Science \& Business Media}
}

@incollection{levy2010spectral,
  title={Spectral mesh processing},
  author={L{\'e}vy, Bruno and Zhang, Hao},
  booktitle={ACM SIGGRAPH 2010 Courses},
  pages={1--312},
  year={2010}
}

@article{mejia2017spectral,
  title={Spectral-based mesh segmentation},
  author={Mejia, Daniel and Ruiz-Salguero, Oscar and Cadavid, Carlos A},
  journal={International Journal on Interactive Design and Manufacturing (IJIDeM)},
  volume={11},
  number={3},
  pages={503--514},
  year={2017},
  publisher={Springer}
}

@inproceedings{lescoat2020spectral,
  title={Spectral mesh simplification},
  author={Lescoat, Thibault and Liu, Hsueh-Ti Derek and Thiery, Jean-Marc and Jacobson, Alec and Boubekeur, Tamy and Ovsjanikov, Maks},
  booktitle={Computer Graphics Forum},
  volume={39},
  number={2},
  pages={315--324},
  year={2020},
  organization={Wiley Online Library}
}

@article{song2014mesh,
  title={Mesh saliency via spectral processing},
  author={Song, Ran and Liu, Yonghuai and Martin, Ralph R and Rosin, Paul L},
  journal={ACM Transactions On Graphics (TOG)},
  volume={33},
  number={1},
  pages={1--17},
  year={2014},
  publisher={ACM New York, NY, USA}
}

@inproceedings{vallet2008spectral,
  title={Spectral geometry processing with manifold harmonics},
  author={Vallet, Bruno and L{\'e}vy, Bruno},
  booktitle={Computer Graphics Forum},
  volume={27},
  number={2},
  pages={251--260},
  year={2008},
  organization={Wiley Online Library}
}

@inproceedings{zhang2010spectral,
  title={Spectral mesh processing},
  author={Zhang, Hao and Van Kaick, Oliver and Dyer, Ramsay},
  booktitle={Computer graphics forum},
  volume={29},
  number={6},
  pages={1865--1894},
  year={2010},
  organization={Wiley Online Library}
}

@inproceedings{sun2009concise,
  title={A concise and provably informative multi-scale signature based on heat diffusion},
  author={Sun, Jian and Ovsjanikov, Maks and Guibas, Leonidas},
  booktitle={Computer graphics forum},
  volume={28},
  number={5},
  pages={1383--1392},
  year={2009},
  organization={Wiley Online Library}
}

@inproceedings{liu2017dirac,
  title={A Dirac operator for extrinsic shape analysis},
  author={Liu, Hsueh-Ti Derek and Jacobson, Alec and Crane, Keenan},
  booktitle={Computer Graphics Forum},
  volume={36},
  number={5},
  pages={139--149},
  year={2017},
  organization={Wiley Online Library}
}

@article{whittle1963stochastic,
  title={Stochastic-processes in several dimensions},
  author={Whittle, Peter},
  journal={Bulletin of the International Statistical Institute},
  volume={40},
  number={2},
  pages={974--994},
  year={1963},
  publisher={INT STATISTICAL INSTITUTE 428 PRINSES BEATRIXLAEN, VOORBURG, NETHERLANDS}
}

@book{williams2006gaussian,
  title={Gaussian processes for machine learning},
  author={Williams, Christopher KI and Rasmussen, Carl Edward},
  volume={2},
  number={3},
  year={2006},
  publisher={MIT press Cambridge, MA}
}

@article{gillan2025discrete,
  title={Discrete Gaussian Vector Fields On Meshes},
  author={Gillan, Michael and Siegert, Stefan and Youngman, Ben},
  journal={arXiv preprint arXiv:2507.20024},
  year={2025}
}

@article{coveney2020gaussian,
  title={Gaussian process manifold interpolation for probabilistic atrial activation maps and uncertain conduction velocity},
  author={Coveney, Sam and Corrado, Cesare and Roney, Caroline H and O’Hare, Daniel and Williams, Steven E and O’Neill, Mark D and Niederer, Steven A and Clayton, Richard H and Oakley, Jeremy E and Wilkinson, Richard D},
  journal={Philosophical Transactions of the Royal Society A},
  volume={378},
  number={2173},
  pages={20190345},
  year={2020},
  publisher={The Royal Society Publishing}
}

@inproceedings{borovitskiy2021matern,
  title={Mat{\'e}rn Gaussian processes on graphs},
  author={Borovitskiy, Viacheslav and Azangulov, Iskander and Terenin, Alexander and Mostowsky, Peter and Deisenroth, Marc and Durrande, Nicolas},
  booktitle={International Conference on Artificial Intelligence and Statistics},
  pages={2593--2601},
  year={2021},
  organization={PMLR}
}

@article{maddix2021modeling,
  title={Modeling Advection on Directed Graphs using Mat$\backslash$'ern Gaussian Processes for Traffic Flow},
  author={Maddix, Danielle C and Saad, Nadim and Wang, Yuyang},
  journal={arXiv preprint arXiv:2201.00001},
  year={2021}
}

@article{bolin2023statistical,
  title={Statistical inference for Gaussian Whittle-Mat$\backslash$'ern fields on metric graphs},
  author={Bolin, David and Simas, Alexandre and Wallin, Jonas},
  journal={arXiv preprint arXiv:2304.10372},
  year={2023}
}

@article{borovitskiy2020matern,
  title={Mat{\'e}rn Gaussian processes on Riemannian manifolds},
  author={Borovitskiy, Viacheslav and Terenin, Alexander and Mostowsky, Peter and others},
  journal={Advances in Neural Information Processing Systems},
  volume={33},
  pages={12426--12437},
  year={2020}
}

@article{rosa2023posterior,
  title={Posterior contraction rates for Mat{\'e}rn Gaussian processes on Riemannian manifolds},
  author={Rosa, Paul and Borovitskiy, Slava and Terenin, Alexander and Rousseau, Judith},
  journal={Advances in Neural Information Processing Systems},
  volume={36},
  pages={34087--34121},
  year={2023}
}

@article{ovsjanikov2012functional,
  title={Functional maps: a flexible representation of maps between shapes},
  author={Ovsjanikov, Maks and Ben-Chen, Mirela and Solomon, Justin and Butscher, Adrian and Guibas, Leonidas},
  journal={ACM Transactions on Graphics (ToG)},
  volume={31},
  number={4},
  pages={1--11},
  year={2012},
  publisher={ACM New York, NY, USA}
}

@inproceedings{gat2025anytop,
  title={Anytop: Character animation diffusion with any topology},
  author={Gat, Inbar and Raab, Sigal and Tevet, Guy and Reshef, Yuval and Bermano, Amit Haim and Cohen-Or, Daniel},
  booktitle={Proceedings of the Special Interest Group on Computer Graphics and Interactive Techniques Conference Conference Papers},
  pages={1--10},
  year={2025}
}

@inproceedings{
tevet2023human,
title={Human Motion Diffusion Model},
author={Guy Tevet and Sigal Raab and Brian Gordon and Yoni Shafir and Daniel Cohen-or and Amit Haim Bermano},
booktitle={The Eleventh International Conference on Learning Representations },
year={2023},
url={https://openreview.net/forum?id=SJ1kSyO2jwu}
}

@inproceedings{shafir2024human,
  title={Human Motion Diffusion as a Generative Prior},
  author={Shafir, Yoni and Tevet, Guy and Kapon, Roy and Bermano, Amit Haim},
  year={2024},
  booktitle={The Twelfth International Conference on Learning Representations}
}

@article{zhong2025sketch2anim,
  title={Sketch2anim: Towards transferring sketch storyboards into 3d animation},
  author={Zhong, Lei and Guo, Chuan and Xie, Yiming and Wang, Jiawei and Li, Changjian},
  journal={ACM Transactions on Graphics (TOG)},
  volume={44},
  number={4},
  pages={1--15},
  year={2025},
  publisher={ACM New York, NY, USA}
}

@article{liu2023meshdiffusion,
  title={Meshdiffusion: Score-based generative 3d mesh modeling},
  author={Liu, Zhen and Feng, Yao and Black, Michael J and Nowrouzezahrai, Derek and Paull, Liam and Liu, Weiyang},
  journal={arXiv preprint arXiv:2303.08133},
  year={2023}
}

@inproceedings{shen2021dmtet,
title = {Deep Marching Tetrahedra: a Hybrid Representation for High-Resolution 3D Shape Synthesis},
author = {Tianchang Shen and Jun Gao and Kangxue Yin and Ming-Yu Liu and Sanja Fidler},
year = {2021},
booktitle = {Advances in Neural Information Processing Systems (NeurIPS)}
}

@inproceedings{ren2024xcube,
  title={Xcube: Large-scale 3d generative modeling using sparse voxel hierarchies},
  author={Ren, Xuanchi and Huang, Jiahui and Zeng, Xiaohui and Museth, Ken and Fidler, Sanja and Williams, Francis},
  booktitle={Proceedings of the IEEE/CVF conference on computer vision and pattern recognition},
  pages={4209--4219},
  year={2024}
}

@article{poole2022dreamfusion,
  title={Dreamfusion: Text-to-3d using 2d diffusion},
  author={Poole, Ben and Jain, Ajay and Barron, Jonathan T and Mildenhall, Ben},
  journal={arXiv preprint arXiv:2209.14988},
  year={2022}
}

@inproceedings{xiang2025structured,
  title={Structured 3d latents for scalable and versatile 3d generation},
  author={Xiang, Jianfeng and Lv, Zelong and Xu, Sicheng and Deng, Yu and Wang, Ruicheng and Zhang, Bowen and Chen, Dong and Tong, Xin and Yang, Jiaolong},
  booktitle={Proceedings of the Computer Vision and Pattern Recognition Conference},
  pages={21469--21480},
  year={2025}
}

@article{zhang2024clay,
  title={Clay: A controllable large-scale generative model for creating high-quality 3d assets},
  author={Zhang, Longwen and Wang, Ziyu and Zhang, Qixuan and Qiu, Qiwei and Pang, Anqi and Jiang, Haoran and Yang, Wei and Xu, Lan and Yu, Jingyi},
  journal={ACM Transactions on Graphics (TOG)},
  volume={43},
  number={4},
  pages={1--20},
  year={2024},
  publisher={ACM New York, NY, USA}
}

@inproceedings{tan2018variational,
  title={Variational autoencoders for deforming 3d mesh models},
  author={Tan, Qingyang and Gao, Lin and Lai, Yu-Kun and Xia, Shihong},
  booktitle={Proceedings of the IEEE conference on computer vision and pattern recognition},
  pages={5841--5850},
  year={2018}
}

@inproceedings{yuan2020mesh,
  title={Mesh variational autoencoders with edge contraction pooling},
  author={Yuan, Yu-Jie and Lai, Yu-Kun and Yang, Jie and Duan, Qi and Fu, Hongbo and Gao, Lin},
  booktitle={Proceedings of the IEEE/CVF conference on computer vision and pattern recognition workshops},
  pages={274--275},
  year={2020}
}

@article{tan2021variational,
  title={Variational autoencoders for localized mesh deformation component analysis},
  author={Tan, Qingyang and Zhang, Ling-Xiao and Yang, Jie and Lai, Yu-Kun and Gao, Lin},
  journal={IEEE Transactions on Pattern Analysis and Machine Intelligence},
  volume={44},
  number={10},
  pages={6297--6310},
  year={2021},
  publisher={IEEE}
}

@inproceedings{muralikrishnan2022glass,
  title={Glass: Geometric latent augmentation for shape spaces},
  author={Muralikrishnan, Sanjeev and Chaudhuri, Siddhartha and Aigerman, Noam and Kim, Vladimir G and Fisher, Matthew and Mitra, Niloy J},
  booktitle={Proceedings of the IEEE/CVF Conference on Computer Vision and Pattern Recognition},
  pages={18552--18561},
  year={2022}
}

@inproceedings{huang2021arapreg,
  title={Arapreg: An as-rigid-as possible regularization loss for learning deformable shape generators},
  author={Huang, Qixing and Huang, Xiangru and Sun, Bo and Zhang, Zaiwei and Jiang, Junfeng and Bajaj, Chandrajit},
  booktitle={Proceedings of the IEEE/CVF international conference on computer vision},
  pages={5815--5825},
  year={2021}
}

@inproceedings{dinh2025geometry,
  title={Geometry in Style: 3D Stylization via Surface Normal Deformation},
  author={Dinh, Nam Anh and Lang, Itai and Kim, Hyunwoo and Stein, Oded and Hanocka, Rana},
  booktitle={Proceedings of the Computer Vision and Pattern Recognition Conference},
  pages={28456--28467},
  year={2025}
}

@inproceedings{kim2025meshup,
  title={Meshup: Multi-target mesh deformation via blended score distillation},
  author={Kim, Hyunwoo and Lang, Itai and Aigerman, Noam and Groueix, Thibault and Kim, Vladimir G and Hanocka, Rana},
  booktitle={2025 International Conference on 3D Vision (3DV)},
  pages={222--239},
  year={2025},
  organization={IEEE}
}

@inproceedings{michel2022text2mesh,
  title={Text2mesh: Text-driven neural stylization for meshes},
  author={Michel, Oscar and Bar-On, Roi and Liu, Richard and Benaim, Sagie and Hanocka, Rana},
  booktitle={Proceedings of the IEEE/CVF conference on computer vision and pattern recognition},
  pages={13492--13502},
  year={2022}
}

@inproceedings{gao2023textdeformer,
  title={Textdeformer: Geometry manipulation using text guidance},
  author={Gao, William and Aigerman, Noam and Groueix, Thibault and Kim, Vova and Hanocka, Rana},
  booktitle={ACM SIGGRAPH 2023 conference proceedings},
  pages={1--11},
  year={2023}
}

@article{fumero2020nonlinear,
  title={Nonlinear spectral geometry processing via the TV transform},
  author={Fumero, Marco and M{\"o}ller, Michael and Rodol{\`a}, Emanuele},
  journal={ACM Transactions on Graphics (TOG)},
  volume={39},
  number={6},
  pages={1--16},
  year={2020},
  publisher={ACM New York, NY, USA}
}

@inproceedings{wang2025headevolver,
  title={Headevolver: Text to head avatars via expressive and attribute-preserving mesh deformation},
  author={Wang, Duotun and Meng, Hengyu and Cai, Zeyu and Shao, Zhijing and Liu, Qianxi and Wang, Lin and Fan, Mingming and Zhan, Xiaohang and Wang, Zeyu},
  booktitle={2025 International Conference on 3D Vision (3DV)},
  pages={211--221},
  year={2025},
  organization={IEEE}
}

@article{lu2025ll3m,
  title={Ll3m: Large language 3d modelers},
  author={Lu, Sining and Chen, Guan and Dinh, Nam Anh and Lang, Itai and Holtzman, Ari and Hanocka, Rana},
  journal={arXiv preprint arXiv:2508.08228},
  year={2025}
}

@article{siddiqui2023meshgpt,
  title={MeshGPT: Generating Triangle Meshes with Decoder-Only Transformers},
  author={Siddiqui, Yawar and Alliegro, Antonio and Artemov, Alexey and Tommasi, Tatiana and Sirigatti, Daniele and Rosov, Vladislav and Dai, Angela and Nie{\ss}ner, Matthias},
  journal={arXiv preprint arXiv:2311.15475},
  year={2023}
}

@inproceedings{nash2020polygen,
  title={Polygen: An autoregressive generative model of 3d meshes},
  author={Nash, Charlie and Ganin, Yaroslav and Eslami, SM Ali and Battaglia, Peter},
  booktitle={International conference on machine learning},
  pages={7220--7229},
  year={2020},
  organization={PMLR}
}

@article{gao2019sdm,
  title={SDM-NET: Deep generative network for structured deformable mesh},
  author={Gao, Lin and Yang, Jie and Wu, Tong and Yuan, Yu-Jie and Fu, Hongbo and Lai, Yu-Kun and Zhang, Hao},
  journal={ACM Transactions on Graphics (TOG)},
  volume={38},
  number={6},
  pages={1--15},
  year={2019},
  publisher={ACM New York, NY, USA}
}

@inproceedings{liu2021deepmetahandles,
  title={Deepmetahandles: Learning deformation meta-handles of 3d meshes with biharmonic coordinates},
  author={Liu, Minghua and Sung, Minhyuk and Mech, Radomir and Su, Hao},
  booktitle={Proceedings of the IEEE/CVF Conference on Computer Vision and Pattern Recognition},
  pages={12--21},
  year={2021}
}

@article{guo2022systematic,
  title={A systematic survey on deep generative models for graph generation},
  author={Guo, Xiaojie and Zhao, Liang},
  journal={IEEE Transactions on Pattern Analysis and Machine Intelligence},
  volume={45},
  number={5},
  pages={5370--5390},
  year={2022},
  publisher={IEEE}
}

@article{zhang2023survey,
  title={A survey on graph diffusion models: Generative ai in science for molecule, protein and material},
  author={Zhang, Mengchun and Qamar, Maryam and Kang, Taegoo and Jung, Yuna and Zhang, Chenshuang and Bae, Sung-Ho and Zhang, Chaoning},
  journal={arXiv preprint arXiv:2304.01565},
  year={2023}
}

@inproceedings{gao2022get3d,
title={GET3D: A Generative Model of High Quality 3D Textured Shapes Learned from Images},
author={Jun Gao and Tianchang Shen and Zian Wang and Wenzheng Chen and Kangxue Yin and Daiqing Li and Or Litany and Zan Gojcic and Sanja Fidler},
booktitle={Advances In Neural Information Processing Systems},
year={2022}
}

@inproceedings{yang2019pointflow,
  title={Pointflow: 3d point cloud generation with continuous normalizing flows},
  author={Yang, Guandao and Huang, Xun and Hao, Zekun and Liu, Ming-Yu and Belongie, Serge and Hariharan, Bharath},
  booktitle={Proceedings of the IEEE/CVF international conference on computer vision},
  pages={4541--4550},
  year={2019}
}

@article{vahdat2022lion,
  title={Lion: Latent point diffusion models for 3d shape generation},
  author={Vahdat, Arash and Williams, Francis and Gojcic, Zan and Litany, Or and Fidler, Sanja and Kreis, Karsten and others},
  journal={Advances in Neural Information Processing Systems},
  volume={35},
  pages={10021--10039},
  year={2022}
}

@book{chavel1984eigenvalues,
  title={Eigenvalues in Riemannian geometry},
  author={Chavel, Isaac},
  volume={115},
  year={1984},
  publisher={Academic press}
}

@article{wardetzky2008convergence,
  title={Convergence of the cotangent formula: An overview},
  author={Wardetzky, Max},
  journal={Discrete differential geometry},
  pages={275--286},
  year={2008},
  publisher={Springer}
}

@phdthesis{wardetzky2007discrete,
  title={Discrete differential operators on polyhedral surfaces-convergence and approximation},
  author={Wardetzky, Max},
  year={2007}
}

@inproceedings{yan2025object,
  title={An object is worth 64$\times$ 64 pixels: Generating 3d object via image diffusion},
  author={Yan, Xingguang and Lee, Han-Hung and Wan, Ziyu and Chang, Angel X},
  booktitle={2025 International Conference on 3D Vision (3DV)},
  pages={123--133},
  year={2025},
  organization={IEEE}
}

@article{song2020score,
  title={Score-based generative modeling through stochastic differential equations},
  author={Song, Yang and Sohl-Dickstein, Jascha and Kingma, Diederik P and Kumar, Abhishek and Ermon, Stefano and Poole, Ben},
  journal={arXiv preprint arXiv:2011.13456},
  year={2020}
}

@article{liu2022flow,
  title={Flow straight and fast: Learning to generate and transfer data with rectified flow},
  author={Liu, Xingchao and Gong, Chengyue and Liu, Qiang},
  journal={arXiv preprint arXiv:2209.03003},
  year={2022}
}

@article{de2022riemannian,
  title={Riemannian score-based generative modelling},
  author={De Bortoli, Valentin and Mathieu, Emile and Hutchinson, Michael and Thornton, James and Teh, Yee Whye and Doucet, Arnaud},
  journal={Advances in neural information processing systems},
  volume={35},
  pages={2406--2422},
  year={2022}
}

@article{huang2022riemannian,
  title={Riemannian diffusion models},
  author={Huang, Chin-Wei and Aghajohari, Milad and Bose, Joey and Panangaden, Prakash and Courville, Aaron C},
  journal={Advances in Neural Information Processing Systems},
  volume={35},
  pages={2750--2761},
  year={2022}
}

@article{daviet2025neurally,
  title={Neurally integrated finite elements for differentiable elasticity on evolving domains},
  author={Daviet, Gilles and Shen, Tianchang and Sharp, Nicholas and Levin, David IW},
  journal={ACM Transactions on Graphics},
  volume={44},
  number={2},
  pages={1--17},
  year={2025},
  publisher={ACM New York, NY}
}

@inproceedings{tang2024lgm,
  title={Lgm: Large multi-view gaussian model for high-resolution 3d content creation},
  author={Tang, Jiaxiang and Chen, Zhaoxi and Chen, Xiaokang and Wang, Tengfei and Zeng, Gang and Liu, Ziwei},
  booktitle={European Conference on Computer Vision},
  pages={1--18},
  year={2024},
  organization={Springer}
}

@article{KingmaW19,
  author       = {Diederik P. Kingma and
                  Max Welling},
  title        = {An Introduction to Variational Autoencoders},
  journal      = {Found. Trends Mach. Learn.},
  volume       = {12},
  number       = {4},
  pages        = {307--392},
  year         = {2019},
  url          = {https://doi.org/10.1561/2200000056},
  doi          = {10.1561/2200000056},
  bibsource    = {dblp computer science bibliography, https://dblp.org}
}

@article{bolin2024gaussian,
  title={Gaussian Whittle--Mat{\'e}rn fields on metric graphs},
  author={Bolin, David and Simas, Alexandre B and Wallin, Jonas},
  journal={Bernoulli},
  volume={30},
  number={2},
  pages={1611--1639},
  year={2024},
  publisher={Bernoulli Society for Mathematical Statistics and Probability}
}

@article{matern1960spatial,
  title={Spatial variation. Stochastic models and their application to some problems in forest surveys and other sampling investigations.},
  author={Mat{\'e}rn, Bertil},
  year={1960}
}

@article{liu2020neural,
  title={Neural subdivision},
  author={Liu, Hsueh-Ti Derek and Kim, Vladimir G and Chaudhuri, Siddhartha and Aigerman, Noam and Jacobson, Alec},
  journal={ACM Transactions on Graphics (TOG)},
  volume={39},
  number={4},
  pages={124--1},
  year={2020},
  publisher={ACM New York, NY, USA}
}

@inproceedings{ren2024tiger,
  title={TIGER: Time-varying denoising model for 3D point cloud generation with diffusion process},
  author={Ren, Zhiyuan and Kim, Minchul and Liu, Feng and Liu, Xiaoming},
  booktitle={Proceedings of the IEEE/CVF Conference on Computer Vision and Pattern Recognition},
  pages={9462--9471},
  year={2024}
}

@article{perlin1985image,
  title={An image synthesizer},
  author={Perlin, Ken},
  journal={ACM Siggraph Computer Graphics},
  volume={19},
  number={3},
  pages={287--296},
  year={1985},
  publisher={ACM New York, NY, USA}
}

@inproceedings{Liuflowaudio,
  author       = {Alexander H. Liu and
                  Matthew Le and
                  Apoorv Vyas and
                  Bowen Shi and
                  Andros Tjandra and
                  Wei{-}Ning Hsu},
  title        = {Generative Pre-training for Speech with Flow Matching},
  booktitle    = {The Twelfth International Conference on Learning Representations,
                  {ICLR} 2024, Vienna, Austria, May 7-11, 2024},
  publisher    = {OpenReview.net},
  year         = {2024},
  url          = {https://openreview.net/forum?id=KpoQSgxbKH},
  bibsource    = {dblp computer science bibliography, https://dblp.org}
}

@article{li2025flow,
  title={Flow matching meets biology and life science: a survey},
  author={Li, Zihao and Zeng, Zhichen and Lin, Xiao and Fang, Feihao and Qu, Yanru and Xu, Zhe and Liu, Zhining and Ning, Xuying and Wei, Tianxin and Liu, Ge and others},
  journal={arXiv preprint arXiv:2507.17731},
  year={2025}
}

@inproceedings{dupont22,
  author       = {Emilien Dupont and
                  Yee Whye Teh and
                  Arnaud Doucet},
  editor       = {Gustau Camps{-}Valls and
                  Francisco J. R. Ruiz and
                  Isabel Valera},
  title        = {Generative Models as Distributions of Functions},
  booktitle    = {International Conference on Artificial Intelligence and Statistics,
                  {AISTATS} 2022, 28-30 March 2022, Virtual Event},
  series       = {Proceedings of Machine Learning Research},
  volume       = {151},
  pages        = {2989--3015},
  publisher    = {{PMLR}},
  year         = {2022},
  url          = {https://proceedings.mlr.press/v151/dupont22a.html},
  bibsource    = {dblp computer science bibliography, https://dblp.org}
}

@inproceedings{karni2000spectral,
  title={Spectral compression of mesh geometry},
  author={Karni, Zachi and Gotsman, Craig},
  booktitle={Proceedings of the 27th annual conference on Computer graphics and interactive techniques},
  pages={279--286},
  year={2000}
}

@inproceedings{levy2006laplace,
  title={Laplace-beltrami eigenfunctions towards an algorithm that" understands" geometry},
  author={L{\'e}vy, Bruno},
  booktitle={IEEE International Conference on Shape Modeling and Applications 2006 (SMI'06)},
  pages={13--13},
  year={2006},
  organization={IEEE}
}

@incollection{gotsman2003fundamentals,
  title={Fundamentals of spherical parameterization for 3D meshes},
  author={Gotsman, Craig and Gu, Xianfeng and Sheffer, Alla},
  booktitle={ACM SIGGRAPH 2003 Papers},
  pages={358--363},
  year={2003}
}

@article{ben2005optimality,
  title={On the optimality of spectral compression of mesh data},
  author={Ben-Chen, Mirela and Gotsman, Craig},
  journal={ACM Transactions on Graphics (TOG)},
  volume={24},
  number={1},
  pages={60--80},
  year={2005},
  publisher={ACM New York, NY, USA}
}

@inproceedings{zhuravlev2025denoising,
  title={Denoising functional maps: Diffusion models for shape correspondence},
  author={Zhuravlev, Aleksei and L{\"a}hner, Zorah and Golyanik, Vladislav},
  booktitle={Proceedings of the Computer Vision and Pattern Recognition Conference},
  pages={26899--26909},
  year={2025}
}

@inproceedings{pierson2025diffumatch,
  title={DiffuMatch: Category-Agnostic Spectral Diffusion Priors for Robust Non-rigid Shape Matching},
  author={Pierson, Emery and Li, Lei and Dai, Angela and Ovsjanikov, Maks},
  booktitle={Proceedings of the IEEE/CVF International Conference on Computer Vision},
  pages={5745--5756},
  year={2025}
}

@inproceedings{rimon2025fridu,
  title={FRIDU: Functional Map Refinement with Guided Image Diffusion},
  author={Rimon, Avigail Cohen and Ben-Chen, Mirela and Litany, Or},
  booktitle={Computer Graphics Forum},
  volume={44},
  number={5},
  pages={e70203},
  year={2025},
  organization={Wiley Online Library}
}

@inproceedings{muralikrishnan2024temporal,
  title={Temporal residual jacobians for rig-free motion transfer},
  author={Muralikrishnan, Sanjeev and Dutt, Niladri and Chaudhuri, Siddhartha and Aigerman, Noam and Kim, Vladimir and Fisher, Matthew and Mitra, Niloy J},
  booktitle={European Conference on Computer Vision},
  pages={93--109},
  year={2024},
  organization={Springer}
}

@article{sundararaman2024deformation,
  title={Deformation recovery: Localized learning for detail-preserving deformations},
  author={Sundararaman, Ramana and Donati, Nicolas and Melzi, Simone and Corman, Etienne and Ovsjanikov, Maks},
  journal={ACM Transactions on Graphics (TOG)},
  volume={43},
  number={6},
  pages={1--16},
  year={2024},
  publisher={ACM New York, NY, USA}
}

@article{besnier2024pandas,
  title={PaNDaS: Learnable Deformation Modeling with Localized Control},
  author={Besnier, Thomas and Pierson, Emery and Arguillere, Sylvain and Ovsjanikov, Maks and Daoudi, Mohamed},
  journal={arXiv preprint arXiv:2412.02306},
  year={2024}
}
\clearpage
% Appendix
\clearpage

\appendix

\section{Proof of the Properties from Section~\ref{ss:properties}}
\label{ap:properties_proof} 
\mypar{Property~\ref{item:ind}.} The covariance matrix in Equation~\eqref{eq:poisson_spectrum} is diagonal and hence each coefficient $\coef_i$ is sampled independently.  
\mypar{Property~\ref{item:agnostic}.} Let two meshes $\mesh,\othermesh$ have coefficients $\coef_i^\mesh,\coef_j^\othermesh$ with corresponding distributions $\spectraldist_i^\mesh,\spectraldist^\othermesh_j$ and corresponding eigenvalues $\lambda_i^\mesh,\lambda_j^\othermesh$, respectively, with $\left|\lambda_i^\mesh-\lambda_j^\othermesh\right|<\varepsilon$. Then $$W_2\parr{\spectraldist^\mesh_i,\spectraldist^\othermesh_j}=\parr{\parr{\lambda^\mesh_i+\tau}^{-1}-\parr{\lambda^\othermesh_j+\tau}^{-1}}^2\leq \frac{\varepsilon^2}{\tau^4}= \bigo{\varepsilon^2}$$ as $\varepsilon$ goes to zero, where we used the trivial definition of the 2-Wasserstein distance $W_2$ for univariate normal distributions for the first equality. 
\mypar{Property~\ref{item:decay}.} 
Per the assumed approximation of the continuous spectrum, the eigenvalues obey \emph{Weyl's law}~\cite{chavel1984eigenvalues}, applied to a 2-manifold with area $A$, satisfying the asymptotic relation $\lambda_j \sim \frac{4\pi}{A} j$. 
Therefore, 
$$\variance{\sum_{i=k}^n\coef_i} = \sum_{i=k}^n\frac{1}{\parr{\lambda_i+\tau}^{2}} \sim  \sum_{i=k}^n\frac{1}{\parr{\frac{4\pi }{A} i  +\tau}^2}<\sum_{i=k}^n\frac{A^2}{{i}^2},$$ 
which is a convergent series as $n$ tends to infinity.  Hence, there exists  $k$ s.t. the above expression is smaller than $\varepsilon$, for any $n$ (i.e., any mesh resolution). 

\section{Implementation Details}

\subsection{Shape Deformation Network}

Our shape deformation network consists of five PoissonNet~\cite{maesumi2025poissonnet} blocks. Each block has hidden dimension of 128 and VectorMLPs with three layers. Our network takes the noisy mesh's $xyz$ as input, and condition the source mesh $xyz$ and the timestep $t$ at each block. We follow PoissonNet~\cite{maesumi2025poissonnet} to use the same NJF head~\cite{aigerman2022neural} at the end of the network to predict the flow velocity in Jacobian space, which can be transformed to velocity in vertex space via solving Poisson's equation. We set the batch size to be 32, and use equal weights for conditional flow matching losses in vertex and Jacobian space. During training, we clip the norm of the gradient to be below 1. We set the screening term $\tau=100$ for all the experiments. The timings we report in all the experiments were conducted on a single NVIDIA RTX 4090 GPU. 

\mypar{Single source generation and elastic deformation models.} We train the network for 160k iterations with a learning rate of 0.0005 for the first 40k iterations, and a learning rate of 0.0001 for rest of the iterations. All trainings were conducted on a single NVIDIA H100 GPU. The total training time of a model is sensitive to the resolution of the source mesh. It takes around 42 hours to train the single source generation (source mesh at resolution of 18k faces) and around 36 hours to train the elastic deformation models (source mesh at resolution of 9k/10k faces). 

\mypar{Arbitrary 3D human shapes model}. We train the network for 240k iterations with an initial learning rate of 0.0005, and multi-step decay of rate 0.2 at 40k and 160k iterations. Besides the source mesh $xyz$ and timestep $t$, we provide the predictions from another pre-trained eigenvector predictor as additional conditioning signals to each PoissonNet block. They serve as intrinsic positional encoding on the given source mesh. The eigenvector predictor is trained on SMPL human meshes of random poses and shapes with the same training target (i.e. the normalized eigenvectors of the neutral T-pose SMPL human mesh's Laplacian that correspond to the smallest 64 eigenvalues). This eigenvector predictor can provide consistent positional encoding on arbitrary 3D human-like meshes, including both SMPL meshes in arbitrary poses and shapes, and general humanoid meshes, which plays an important role in the success of this model. We cache Cholespy solvers~\cite{nicolet2021large} for all the source meshes, and move them to the GPU when needed instead of creating them on-the-fly to accelerate the training. We train the model on a single NVIDIA H100 GPU for 80 hours. We apply random global scaling and shifts to the source meshes as data augmentations when training both the eigenvector predictor and the FM model to further help the models generalize on arbitrary sources.

\subsection{Deformation Autoencoder} We made the first attempt to explore the possibility of encoding deformations via an autoencoder in Sec.~\ref{ss:alternatives}. Our encoder has five DiffusionNet~\cite{sharp2022diffusionnet} blocks using spectral size of 128. It encodes the input deformation into a 128-dimensional latent code by applying a global mean at the end of the encoder. The latent code is served as the conditioning signal to a PoissonNet~\cite{maesumi2025poissonnet} decoder, consisting of five blocks as well. The encoder and decoder's hidden dimensions are 128 and 256, respectively. We train the autoencoder on our Octopus dataset with a learning rate of 0.001 and batch size of 16 for 40k iterations. 

\section{Datasets}

\subsection{SMPL datasets}
We create datasets of deformed SMPL human meshes in yoga poses from the MOYO dataset~\cite{tripathi20233d}. We generate a dataset of 64k deformed human meshes for the single source generation model and the eigenvector predictor used in arbitrary human source model, and 32k pairs of deformed human meshes as source-target pairs for the arbitrary human source model. We follow~\cite{maesumi2025poissonnet} to sample diverse deformed poses from MOYO dataset's motion captures via a greedy farthest point sampling scheme and generate human bodies using the SMPL-X~\cite{SMPL-X:2019} parametric model. We keep the hands to always be in neutral poses in our datasets. The deformed shapes used in single source generation model are kept to be in a neutral shape, while in arbitrary human source model the body shape parameters are randomly sampled from a Gaussian distribution with a standard deviation of 5 to obtain diverse human bodies. All the meshes in the dataset are at resolution of 18k faces.

\subsection{Elastic deformations datasets}
We apply a physical simulation method, IPC~\cite{li2020incremental}, to simulate the resting states of elastic object on the floor. We downsample our high-resolution mesh assets and convert them to tetrahedral meshes via fTetWild~\cite{hu2020fast}. We simulate object dropping with random initial orientation, as well as initial angular speed randomly sampled between $\pi$ and $3\pi$. All objects have a Young's modulus, Poisson's ratio, and density of 1e4, 0.4, and 1e3, respectively. The simulation runs till the object reaches an equilibrium state on the floor and we collect it as one deformed sample in the training dataset. We convert the resting states back to triangle meshes. To remove the global rotation of the object around the y-axis (up-direction in our coordinate system), we apply PCA on each shape's $x$ and $z$ coordinates, and rotate the mesh such that its largest principal component aligns with the largest component of the source mesh (the canonical template mesh used in simulation). We create datasets for four objects: Octopus, Elephant, Fish, and the Stanford Bunny, and for each object we sample 20k diverse resting states (via farthest point sampling) from 30k post-PCA simulation results. The original Octopus, Elephant, and Stanford Bunny meshes have 100k, 400k, and 70k faces, respectively. The low resolution meshes used to train our models for Octopus, Elephant, and Stanford Bunny have 9k, 10k, and 9k faces, respectively. We use Fish in original resolution (7k faces) to train our model.

\subsection{GMM datasets}

We follow the setting from MDF~\cite{elhag2023manifold} to create datasets of random Gaussian Mixture Models (GMM) on a uniformly triangulated Stanford Bunny of 70k faces. We randomly sample GMM of three non-overlapping centers on the mesh. Note that the probability of sampling a vertex as Gaussian center is weighted by its vertex mass, to ensure uniform distribution of Gaussians on the surface. We create a dataset of 32k random GMMs as the training dataset. We also create a test mesh by downsampling the left half of the bunny, resulting in a mesh of non-uniform triangulation (40k faces). We generate 1k random GMMs on the test mesh using the same sampling strategy, yielding a reference set for evaluation.

\section{Evaluation Metrics}

\mypar{MMD and COV.} As initially proposed by~\cite{achlioptas2018learning}, MMD reports the average distance between generated samples and their closest samples in the reference set, providing a measure of fidelity. On the other hand, COV computes the percentage of the samples in the reference set being closest sample to the generated data. These two metrics are complementary and together they represent the quality of the generation. We use average $l$2 distance weighted by the corresponding vertex mass as the distance metrics for MMD and COV.

\mypar{SMPL deformation metrics.} To measure how realistic the generated deformations are, we introduce a new metric for quantitatively evaluating the generated SMPL shapes, by taking advantage of the differentiable SMPL-X~\cite{SMPL-X:2019} parametric model. We regress the SMPL pose parameters that best match the generated shape by gradient descent. We then use the SMPL-X model and the obtained pose parameters to generate a deformed mesh. We measure the mean squared error (MSE) weighted by the vertex mass as our metric. This metric is more accurate than the MMD in terms of measuring how close the generated samples are to the real deformation space of human. 

\begin{table}[t]
\centering
\caption{Comparison with baseline methods.}
\label{tab:tiger}
\begin{tabular}{l cc cc cc}
\toprule
\multirow{2}{*}{\textbf{Method}}  & \multicolumn{2}{c}{\textbf{MMD} ($\times10^{-2}$) $\downarrow$} & \multicolumn{2}{c}{\textbf{COV} $\uparrow$ } & \multicolumn{2}{c}{\textbf{1-NNA} $\downarrow$} \\
\cmidrule(lr){2-3} \cmidrule(lr){4-5} \cmidrule(lr){6-7}
 & CD & EMD & CD & EMD & CD & EMD \\
\midrule
TIGER  & \textbf{0.56} & \textbf{2.53} & 0.456 & 0.456 & \textbf{0.595} & \textbf{0.635} \\
\textbf{Ours}  & 0.89 & 3.05 & \textbf{0.490} & \textbf{0.510} & 0.694 & 0.669 \\
\bottomrule
\end{tabular}
\end{table}

\begin{figure}[t]       
    \centering      \includegraphics[width=\linewidth]{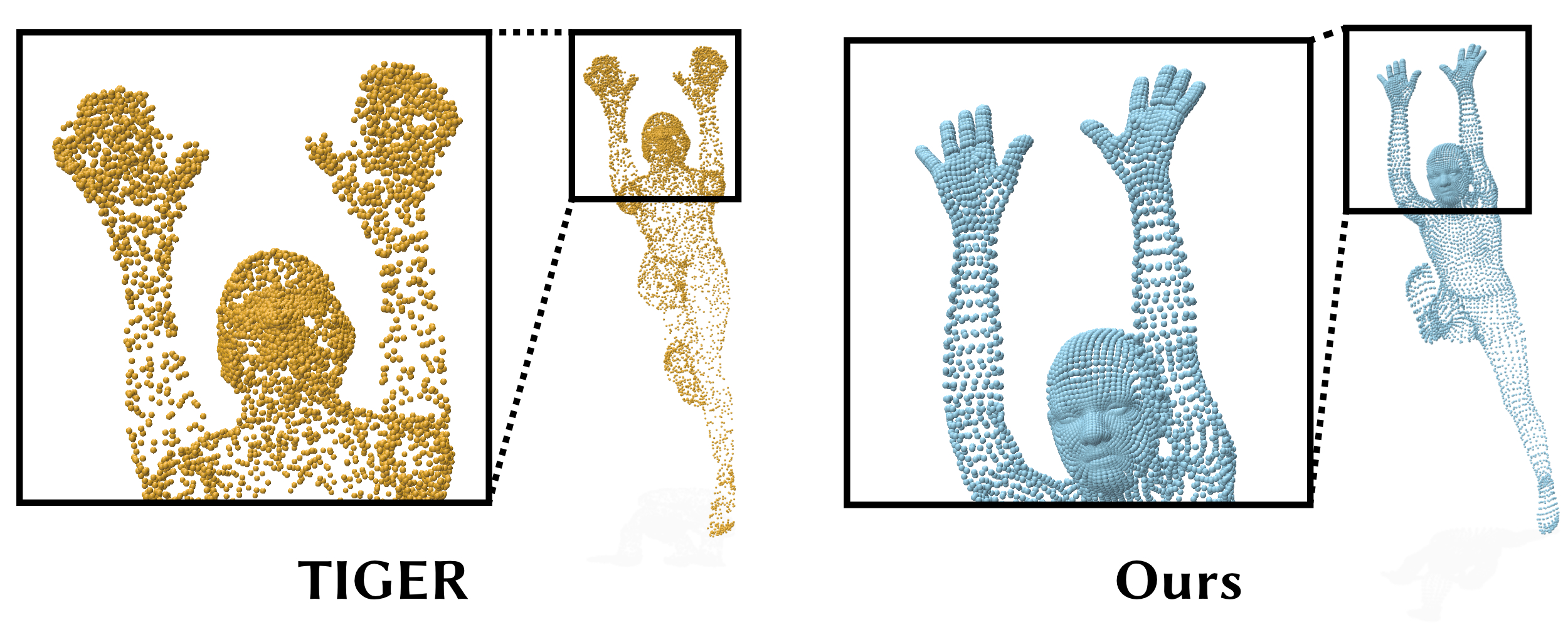}
    \caption{\textbf{Comparison to  state of the art in point cloud generation, TIGER~\cite{ren2024tiger}.} By using a triangulation-agnostic intrinsic network over a mesh, we are able to preserve and produce minute details such as the face and individual fingers; these are completely gone from the point cloud generated by TIGER.
    \label{fig:tiger}  
    }
\end{figure}

\section{Comparison to Point Cloud Generation}
\label{s:TIGER}

Different 3D representations serve different goals (e.g., voxels are better fit for volumetric data), and hence generative methods on meshes cannot be compared directly to other representations, as their goal is to produce details on the vertices of an existing mesh. To further display this point, we evaluate the differences between our mesh-based generative method and a point cloud generative method,  TIGER~\cite{ren2024tiger}, which is currently the state-of-the-art diffusion-based method that applies noising and denoising on point clouds directly. We compare the two on the experiment from Section~\ref{sss:singleSMPL}.  We use the vertices of  the SMPL human deformation dataset as point clouds, and train TIGER from scratch on our dataset to generate deformed SMPL human as point clouds. 

We then compare to our method by treating its generated samples as point clouds as well.  We use the commonly used metrics for point cloud generation (MMD, COV, and 1-NNA) under two different distance metrics, Chamfer distance (CD) and Earth Mover's Distance (EMD) (see TIGER~\cite{ren2024tiger}). Table~\ref{tab:tiger} shows that TIGER slightly outperforms us on most of these metrics, implying better adherence to the target distribution. These results make sense as each method aims for a different task; namely, our method needs to also account for the mesh connectivity and not just the points. 

The gain from using a mesh as opposed to a point cloud is critical, and can be appreciated in Figure~\ref{fig:tiger}: while TIGER~\cite{ren2024tiger} generates a point cloud that well-minimizes the chamfer distance to a valid SMPL human, the lack of connectivity information leads it to completely omit small geometric details (which chamfer distance is insensitive to and hence does not quantify): separate fingers with gaps are non-existent, and the model has a head but no face. This is in contrast to the point cloud from our mesh-based network, which clearly exhibits fingers and a face, due to our method utilizing a triangulation-agnostic framework which operates on the mesh intrinsically  and thereby preserves minute detail.

\section{FEM Discretization}
\label{s:fem}

Given a signal $\signal$, we have values $\signal_i$ on the vertices; we interpret these values in standard linear finite-elements fashion, as representing a piecewise-linear function $\signal(p)$ defined over the triangles as the unique function that is linear on each triangle and interpolates the values $\signal_i$ on the vertices. Given another signal $\textbf{g}$, the inner product between the two piecewise-linear functions is defined as the integral over the domain $\Omega$:

\begin{equation}
\langle f, g \rangle_{L^2} = \int_{\Omega} f(p)g(p) \, dA.
\end{equation}
Substituting the interpolants $f(p) = \sum_i f_i \phi_i(p)$ and $g(p) = \sum_j g_j \phi_j(p)$, where $\{\phi_i\}$ are the standard Lagrange "hat" basis functions associated with each vertex, the inner product is expressed in matrix form as $\mathbf{f}^\top \mathbf{M} \mathbf{g}$. $\mathbf{M} \in \mathbb{R}^{n \times n}$ is the (consistent) \textit{mass matrix}, which is a symmetric matrix with entries:
\begin{equation}
\mass_{ij} = \int_{\Omega} \phi_i(p) \phi_j(p) \, dA.
\end{equation}

To simplify the metric and diagonalize the system, the \textit{lumped} mass matrix $\mathbf{M}$ is frequently employed. This diagonal matrix is defined by summing the rows of the consistent mass matrix:
\begin{equation}
(\mathbf{M})_{ii} = \int_{\Omega} \phi_i(p) \, dA.
\end{equation}
Geometrically, $(\mathbf{M})_{ii}$ corresponds to the area of the dual cell associated with vertex $i$. In practice, this is implemented as the barycentric area (one-third of the area of all incident triangles).

The cotangent Laplacian $\mathbf{L} \in \mathbb{R}^{n \times n}$ is often referred to as the stiffness matrix, and represents the discretization of the Dirichlet energy $\frac{1}{2}\int_{\Omega} \|\nabla f\|^2 \, dA$. Its entries are given by:
\begin{equation}
L_{ij} = \int_{\Omega} \nabla \phi_i(p) \cdot \nabla \phi_j(p) \, dA.
\end{equation}
For a triangle mesh, this yields the standard cotangent weight formula for $i \neq j$:
\begin{equation}
L_{ij} = \frac{1}{2}(\cot \alpha_{ij} + \cot \beta_{ij}),
\end{equation}
where $\alpha_{ij}$ and $\beta_{ij}$ are the angles opposite the edge $(i, j)$. The diagonal entries are defined such that rows sum to zero: $L_{ii} = -\sum_{j \in \mathcal{N}(i)} L_{ij}$, where $\mathcal{N}(i)$ is the neighborhood of $i$ (excluding $i$ itself). \pagebreak

\end{document}